\documentclass[]{IEEEtran}
\IEEEoverridecommandlockouts
\ifCLASSINFOpdf
  \usepackage[pdftex]{graphicx}
\else
   \usepackage[dvips]{graphicx}
\fi
\usepackage[ruled,vlined]{algorithm2e} 
\usepackage[dvipsnames]{xcolor}
\usepackage{soul,framed}
\usepackage[cmex10]{amsmath}
\usepackage{flushend}
\usepackage{cite}
\usepackage[squaren,Gray]{SIunits}
\usepackage{float}
\usepackage{comment}
\usepackage[noend]{algpseudocode}
\usepackage[utf8]{inputenc}
\usepackage{tikz}

\usepackage{stfloats}
\usepackage{amssymb}

\ifCLASSOPTIONcompsoc
    \usepackage[caption=false, font=normalsize, labelfont=sf, textfont=sf]{subfig}
\else
    \usepackage[caption=false, font=footnotesize]{subfig}
\fi

\newtheorem{prop}{Proposition}
\colorlet{shadecolor}{yellow}

\begin{document}

\title{RIS-Aided Covert Communications with Statistical CSI: A Multiport Network Theory Approach
}

\author{Andrea~Abrardo,~\IEEEmembership{Senior Member,~IEEE} and Giulio~Bartoli,~\IEEEmembership{Member,~IEEE}
\thanks{A.~Abrardo and G.~Bartoli are with the Department of Information Engineering and Mathematics at the University of Siena, Siena, Italy.
The work of A.~Abrardo is supported by the 6G SHINE and by the 5G COMPAD - European Defence Fund (EDF)-2021-C4ISR-D european projects. The work of G.~Bartoli is supported by the PSR 2023 - New Frontiers and by the European fund FSE REACT-EU, PON Ricerca e Innovazione 2014-2020.}
}

\maketitle

\begin{abstract}
A novel framework for covert communications aided by Reconfigurable Intelligent Surfaces (RIS) is proposed.
In this general framework, the use of multiport network theory for modelling the RIS consider various aspects that traditional RIS models in communication theory often overlook, including mutual coupling between elements and the impact of structural scattering.
Moreover, the transmitter has only limited knowledge about the channels of the warden and the intended receiver.
The proposed approach is validated through numerical results, demonstrating that communication with the legitimate user is successfully achieved while satisfying the covertness constraint.
\end{abstract}


\section{Introduction}\label{Intro}
In our hyper-connected era, the demand for secure and covert communication has intensified.
Covert communications \cite{CovertComm}, also known as clandestine or stealthy techniques, are crucial in modern systems, particularly in military contexts.
These methods enable the discreet exchange of sensitive information, protecting it from adversaries, classically referred to as wardens, in both physical and cyber domains.

In national security, there is a growing trend to minimize electronic emissions to maintain operational security \cite{angevine2019}.
Emitted signals can be intercepted, revealing sensitive information and compromising security.
As a result, devices like mobile phones are increasingly viewed as risks, leading to greater caution among individuals and organizations.
This shift underscores the importance of managing and suppressing electronic signals to protect sensitive information and ensure operational integrity.
To address these challenges, researchers have shifted their focus toward innovative solutions, prioritizing the use of emerging technologies to enhance the covert communication capabilities of security-sensitive organizations.

One such technology that has gained considerable attention in recent years is Reconfigurable Intelligent Surfaces (RIS), a concept that has the potential to significantly impact wireless communication systems by enabling dynamic manipulation of electromagnetic waves \cite{RIS}.
By deploying a network of intelligent reflecting elements, RIS enables precise control over signal propagation, allowing for enhanced coverage, increased spectral efficiency, and improved reliability in communication links.
In the military context, the potential applications of RIS in covert communications are manifold, offering opportunities to augment operational security, mitigate electronic warfare threats, and facilitate clandestine information exchange in hostile environments.

\subsection{State of the art and contribution}
The goal of covert communications is to enable wireless communication between two users while minimizing the risk of detection by an eavesdropper.
The seminal work in \cite{CovertCommLimits} established the fundamental limit for covert communication, sparking renewed research interest.
In recent years, researchers have explored covert communication under various scenarios, addressing practical constraints such as achieving an error-free privacy rate without detection using multiple input-multiple output (MIMO) techniques \cite{UndetectableComm}, exploiting the presence of jammers \cite{CovertCommUninformedJammer}, and incorporating noise uncertainty to analyze covert effectiveness statistically \cite{CovertCommNoiseUncertainty}.
Studies have also demonstrated covert communication in relay networks \cite{CovertCommRelay} and investigated the use of artificial intelligence methodologies \cite{AI4CovertComm}.

Reconfigurable Intelligent Surfaces (RIS), also known as Intelligent Reflective Surfaces (IRS), have become a rapidly growing research area over the past five years and have the potential to be a transformative technology for enhancing covert communications.
RIS are utilized for improving signal quality in SISO systems \cite{SISORIS}, optimizing sum-rate performance in multi-user networks \cite{MURIS}, increasing energy efficiency \cite{EnergyEfficiencyRIS}, enabling user localization and tracking \cite{LocalizationRIS}, and enhancing transmission security \cite{SecrecyRIS}.
Recent research has focused on refining electromagnetic models of RIS \cite{RisScatteringParameterNetworkAnalysis,abrardo_E}, revealing key features like element coupling, phase shift-gain interdependence, and the presence of specular components—crucial for signal concealment.
These characteristics can be effectively modeled using a multiport (MP) network model, initially introduced in \cite{GradoniEnd2End} for RIS-aided channels.
This nonlinear model has since been applied to various scenarios, including SISO, MIMO, and different channel models \cite{DR2,ABR_MUTUAL}.

Nevertheless, the research on RIS-aided covert communications \cite{IdeaRISCC} is still limited, with only a few studies exploring this area.
The role of RIS in enhancing the detection capability of legitimate receivers while limiting the eavesdropper's listening ability is examined in \cite{CovertDetectionRIS} and \cite{MIMORISCC}.
Energy efficiency in RIS-aided covert communications is analyzed in \cite{EnergyEfficiencyRISCC1,EnergyEfficiencyRISCC2}, and the use of RIS alongside a friendly jammer is explored in \cite{FriendlyJammerRISCC}.
Few studies have addressed scenarios with limited information about the warden's channel, which is the most typical situation in real-world contexts, as the warden is generally an adversary outside of the network.
The case where only the first-order statistical information, i.e., mean and variance, of the Alice-to-Willie and RIS-to-Willie channels are available, is considered in \cite{CovertCommRIS,WardenStatCsiRISCC,DelayContraintRISCC}.
In \cite{CovertCommRIS} and \cite{WardenStatCsiRISCC}, transmission power and RIS configuration are optimized for an instantaneous detection model of Willie.
The study in \cite{DelayContraintRISCC} focuses on the case where Willie has a time window to observe the signal and make his decision.
In \cite{ActivePassiveBeamFormingRISCC}, channel second-order statistic is exploited so that joint precoder and RIS can be optimized to maximize covert rate.
Here the direct path between Alice and Bob is blocked, and Bob transmits a jamming signal for covertness purposes.
Moreover, all the channels are Rayleigh distributed, i.e., without line of sight.
In all the aforementioned works, the customary assumption is made that the RIS consists of ideal scatterers, with possible attenuation independent of the phase shift.

In light of the state of the art of covert communications supported by RIS, the innovative contributions of the paper are summarized below:
\begin{itemize}
    \item We propose a framework to optimize covert communication performance with RIS, extending existing literature to cases where the transmitter has knowledge of the second-order channel statistics for both the warden and the legitimate receiver, considering links with both LOS (Line of Sight) and NLOS (Non-Line of Sight) conditions.
    \item We propose an optimized transmission scheme for Alice with a covertness constraint, separating precoder/RIS configuration from power allocation.
    We show that this approach is optimal for Willie's standard decision strategy, and suboptimal for more complex strategies.
    \item We consider a generalized RIS model that extends the traditional communications model by incorporating critical electromagnetic phenomena, such as structural scattering, coupling between RIS elements, and the interdependence between phase shift and attenuation, which significantly impact performance.
    \item We consider a practical scenario where Alice leverages second-order channel statistics, derived from the nodes' positions and the channel model.
    Our results show that using second-order statistics significantly improves performance compared to the baseline case, which only uses first-order statistics \cite{CovertCommRIS,WardenStatCsiRISCC,DelayContraintRISCC}.
    These results also highlight that neglecting the structural component can severely degrade performance in certain situations.
    Moreover, a peculiarity of covert communications is demonstrated. In particular, while in classical communication scenarios it is known that if the direct link between the transmitter and receiver is sufficiently good, the RIS has little to no utility, in covert communications, even in the presence of strong direct paths, the use of the RIS can provide significant benefits. This holds true even with partial knowledge of the CSI.
\end{itemize}

\subsection{Paper Outline and Notation}
The rest of the paper is organized as follow: Section~\ref{sec:SystemModel} presents the system model, Section~\ref{sec:ProblemFormulation} provides the formalization of the problem and Section~\ref{sec:RisOptimization} discusses the solution, whose results are analyzed in Section~\ref{sec:NumericalResults}.
Finally, conclusions are drawn in Section~\ref{sec:Conclusion}.

{\sl Notation}: Unless otherwise specified, matrices are denoted by bold uppercase letters (i.e., $\mathbf{X}$), vectors are represented by bold lowercase letters (i.e., $\mathbf{x}$), and scalars are denoted by normal font (i.e., $x$). $(\cdot)^{\mathrm{T}}$, $(\cdot)^{\mathrm{H}}$ and $(\cdot)^{-1}$ stand for the transpose, Hermitian transpose and inverse of the matrices. The symbol $\odot$ represents the Hadamard (element-wise) product while with $\text{diag}\left(\mathbf{x}\right)$ we mean the diagonal matrix obtained from the element of vector $\mathbf{x}$, while with $\text{diag}\left(\mathbf{X}\right)$ we mean the vector obtained from the elements of the main diagonal of matrix $\mathbf{X}$. The notation $||\mathbf{x}||$ signifies the Euclidean norm of the vector $\mathbf{x}$, $\left\|\mathbf{X}\right\|$ is the Frobenius norm of the matrix $\mathbf{X}$, and $\mathbb{E}\{\cdot\}$ represents the statistical expectation.

\section{System Model}\label{sec:SystemModel}
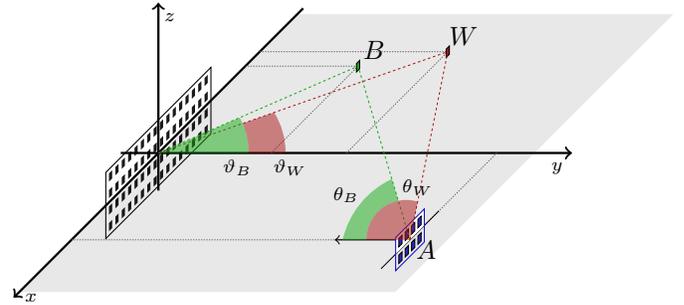
\begin{figure}[ht]
    \centering
        \newlength{\AliceZeroX}
        \setlength{\AliceZeroX}{3.0pt}
        \newlength{\AliceZeroY}
        \setlength{\AliceZeroY}{4.5pt}
        
        \newlength{\BobZeroX}
        \setlength{\BobZeroX}{-3.0pt}
        \newlength{\BobZeroY}
        \setlength{\BobZeroY}{1.5pt}
        
        \newlength{\WillieZeroX}
        \setlength{\WillieZeroX}{-3.5pt}
        \newlength{\WillieZeroY}
        \setlength{\WillieZeroY}{2.5pt}
        
        \newlength{\DipSize}
        \setlength{\DipSize}{0.11pt}
        
        \newlength{\AliceInitX}
        \setlength{\AliceInitX}{\dimexpr \AliceZeroX - 3.5\DipSize \relax}
        \newlength{\AliceInitZ}
        \setlength{\AliceInitZ}{\dimexpr \DipSize/2}
        
        \newlength{\RisInitX}
        \setlength{\RisInitX}{\dimexpr - 15.5\DipSize \relax}
        \newlength{\RisInitZ}
        \setlength{\RisInitZ}{\dimexpr -3.5\DipSize}
        
        \definecolor{PositionColor}{RGB}{100,100,100}
        \definecolor{AliceColor}{RGB}{20,20,200}
        \definecolor{BobColor}{RGB}{20,170,20}
        \definecolor{WillieColor}{RGB}{170,20,20}
        
        \begin{tikzpicture}
        
          \fill[gray!20,opacity=0.9] (0,0,-4.8) -- (5,0,-4.8) -- (5,0,4.8) -- (0,0,4.8) -- cycle;
        
          \draw[->,line width=0.3mm] (-0.5,0,0) -- (5.5,0,0);
          \node[] (y) at (5.5-0.2,-0.2,0) {$\scriptstyle y$};
          \draw[->,line width=0.3mm] (0,-0.5,0) -- (0,2.0,0);
          \node[] (z) at (0.15,2.0-0.2,0) {$\scriptstyle z$};
          \draw[->,line width=0.3mm] (0,0,-5.0) -- (0,0,5);
          \node[] (x) at (0.23,0,5) {$\scriptstyle x$};
        
          \node[] (A) at (\AliceZeroY+0.35,0.0,\AliceZeroX+0.35) {$A$};
          \draw[PositionColor,dash pattern=on 0.5pt off 0.5pt,line width=0.08mm] (0,0,\AliceZeroX) -- (\AliceZeroY,0,\AliceZeroX) node[anchor=north east] {};
          \draw[PositionColor,dash pattern=on 0.5pt off 0.5pt,line width=0.08mm] (\AliceZeroY,0,0) -- (\AliceZeroY,0,\AliceZeroX) node[anchor=south east] {};
          \draw[-,line width=0.1mm] (\AliceZeroY,0,\AliceZeroX-1.0) -- (\AliceZeroY,0,\AliceZeroX+1.0) node[anchor=north east] {};
          \draw[->,line width=0.15mm] (\AliceZeroY,0,\AliceZeroX) -- (\AliceZeroY-1.0,0,\AliceZeroX) node[anchor=south east] {};
          \fill[white] (\AliceZeroY,0,\AliceInitX-\DipSize) -- (\AliceZeroY,\AliceInitZ+1.5*\DipSize,\AliceInitX-\DipSize) -- (\AliceZeroY,\AliceInitZ+1.5*\DipSize,\AliceInitX+8*\DipSize) -- (\AliceZeroY,0,\AliceInitX+8*\DipSize) -- cycle;
          \draw[AliceColor] (\AliceZeroY,-\AliceInitZ-1.5*\DipSize,\AliceInitX-\DipSize) -- (\AliceZeroY,\AliceInitZ+1.5*\DipSize,\AliceInitX-\DipSize) -- (\AliceZeroY,\AliceInitZ+1.5*\DipSize,\AliceInitX+8*\DipSize) -- (\AliceZeroY,-\AliceInitZ-1.5*\DipSize,\AliceInitX+8*\DipSize) -- cycle;
          \foreach \nn in {0,1,...,3}
          {
            \fill[AliceColor,opacity=0.9] (\AliceZeroY,\AliceInitZ,\AliceInitX+\nn*2*\DipSize) -- (\AliceZeroY,\AliceInitZ+\DipSize,\AliceInitX+\nn*2*\DipSize) -- (\AliceZeroY,\AliceInitZ+\DipSize,\AliceInitX+\nn*2*\DipSize+\DipSize) -- (\AliceZeroY,\AliceInitZ,\AliceInitX+\nn*2*\DipSize+\DipSize) -- cycle;
            \fill[AliceColor,opacity=0.9] (\AliceZeroY,-\AliceInitZ,\AliceInitX+\nn*2*\DipSize) -- (\AliceZeroY,-\AliceInitZ-\DipSize,\AliceInitX+\nn*2*\DipSize) -- (\AliceZeroY,-\AliceInitZ-\DipSize,\AliceInitX+\nn*2*\DipSize+\DipSize) -- (\AliceZeroY,-\AliceInitZ,\AliceInitX+\nn*2*\DipSize+\DipSize) -- cycle;
            \draw[black] (\AliceZeroY,\AliceInitZ,\AliceInitX+\nn*2*\DipSize) -- (\AliceZeroY,\AliceInitZ+\DipSize,\AliceInitX+\nn*2*\DipSize) -- (\AliceZeroY,\AliceInitZ+\DipSize,\AliceInitX+\nn*2*\DipSize+\DipSize) -- (\AliceZeroY,\AliceInitZ,\AliceInitX+\nn*2*\DipSize+\DipSize) -- cycle;
            \draw[black] (\AliceZeroY,-\AliceInitZ,\AliceInitX+\nn*2*\DipSize) -- (\AliceZeroY,-\AliceInitZ-\DipSize,\AliceInitX+\nn*2*\DipSize) -- (\AliceZeroY,-\AliceInitZ-\DipSize,\AliceInitX+\nn*2*\DipSize+\DipSize) -- (\AliceZeroY,-\AliceInitZ,\AliceInitX+\nn*2*\DipSize+\DipSize) -- cycle;
        }
        
          \draw[black] (0,\RisInitZ-\DipSize/2,\RisInitX-\DipSize) -- (0,\RisInitZ+7.5*\DipSize,\RisInitX-\DipSize) -- (0,\RisInitZ+7.5*\DipSize,\RisInitX+32*\DipSize) -- (0,\RisInitZ-\DipSize/2,\RisInitX+32*\DipSize) -- cycle;
          \foreach \nn in {0,1,...,15}
          {
            \foreach \mm in {0,1,...,3}
            {
              \fill[black,opacity=0.9] (0,\RisInitZ+\mm*2*\DipSize,\RisInitX+\nn*2*\DipSize) -- (0,\RisInitZ+\mm*2*\DipSize+\DipSize,\RisInitX+\nn*2*\DipSize) -- (0,\RisInitZ+\mm*2*\DipSize+\DipSize,\RisInitX+\nn*2*\DipSize+\DipSize) -- (0,\RisInitZ+\mm*2*\DipSize,\RisInitX+\nn*2*\DipSize+\DipSize) -- cycle;
            }
          }
        
          \node[] (B) at (\BobZeroY,0.0,\BobZeroX-0.55) {$B$};
          \draw[PositionColor,dash pattern=on 0.5pt off 0.5pt,line width=0.08mm] (0,0,\BobZeroX) -- (\BobZeroY,0,\BobZeroX) node[anchor=north east] {};
          \draw[PositionColor,dash pattern=on 0.5pt off 0.5pt,line width=0.08mm] (\BobZeroY,0,0) -- (\BobZeroY,0,\BobZeroX) node[anchor=south east] {};
          \fill[BobColor] (\BobZeroY,-0.5*\DipSize,\BobZeroX-0.5*\DipSize) -- (\BobZeroY,-0.5*\DipSize,\BobZeroX+0.5*\DipSize) -- (\BobZeroY,+0.5*\DipSize,\BobZeroX+0.5*\DipSize) -- (\BobZeroY,+0.5*\DipSize,\BobZeroX-0.5*\DipSize) -- cycle;
          \draw[black] (\BobZeroY,-0.5*\DipSize,\BobZeroX-0.5*\DipSize) -- (\BobZeroY,-0.5*\DipSize,\BobZeroX+0.5*\DipSize) -- (\BobZeroY,+0.5*\DipSize,\BobZeroX+0.5*\DipSize) -- (\BobZeroY,+0.5*\DipSize,\BobZeroX-0.5*\DipSize) -- cycle;
        
          \node[] (W) at (\WillieZeroY,0.0,\WillieZeroX-0.55) {$W$};
          \draw[PositionColor,dash pattern=on 0.5pt off 0.5pt,line width=0.08mm] (0,0,\WillieZeroX) -- (\WillieZeroY,0,\WillieZeroX) node[anchor=north east] {};
          \draw[PositionColor,dash pattern=on 0.5pt off 0.5pt,line width=0.08mm] (\WillieZeroY,0,0) -- (\WillieZeroY,0,\WillieZeroX) node[anchor=south east] {};
          \fill[WillieColor] (\WillieZeroY,-0.5*\DipSize,\WillieZeroX-0.5*\DipSize) -- (\WillieZeroY,-0.5*\DipSize,\WillieZeroX+0.5*\DipSize) -- (\WillieZeroY,+0.5*\DipSize,\WillieZeroX+0.5*\DipSize) -- (\WillieZeroY,+0.5*\DipSize,\WillieZeroX-0.5*\DipSize) -- cycle;
          \draw[black] (\WillieZeroY,-0.5*\DipSize,\WillieZeroX-0.5*\DipSize) -- (\WillieZeroY,-0.5*\DipSize,\WillieZeroX+0.5*\DipSize) -- (\WillieZeroY,+0.5*\DipSize,\WillieZeroX+0.5*\DipSize) -- (\WillieZeroY,+0.5*\DipSize,\WillieZeroX-0.5*\DipSize) -- cycle;
        
          \draw[BobColor,dash pattern=on 1.0pt off 1.0pt,line width=0.09mm] (0,0,0) -- (\BobZeroY,0,\BobZeroX) node[anchor=south east] {};
          \draw[WillieColor,dash pattern=on 1.0pt off 1.0pt,line width=0.09mm] (0,0,0) -- (\WillieZeroY,0,\WillieZeroX) node[anchor=south east] {};
          \fill[BobColor, opacity=0.5] 
          (\BobZeroY*0.4,0,\BobZeroX*0.4) -- (0,0,0) -- (\BobZeroY*0.8,0,0) arc[start angle=0, end angle=28, radius=1];
          \node[] (varthetaB) at (\BobZeroY*0.7,-0.19pt,0) {$\scriptstyle\vartheta_B$};
          \fill[WillieColor, opacity=0.5] (\BobZeroY*0.8,0,0) arc[start angle=0, end angle=23, radius=1] --(\WillieZeroY*0.3,0,\WillieZeroX*0.3) -- (\WillieZeroY*0.4,0,\WillieZeroX*0.4) arc[start angle=32, end angle=0, radius=1] -- (\WillieZeroY*0.65,0,0);
          \node[] (varthetaW) at (\WillieZeroY*0.7,-0.19pt,0) {$\scriptstyle\vartheta_W$};
          \draw[BobColor,dash pattern=on 1.0pt off 1.0pt,line width=0.09mm] (\AliceZeroY,0,\AliceZeroX) -- (\BobZeroY,0,\BobZeroX) node[anchor=south east] {};
          \draw[WillieColor,dash pattern=on 1.0pt off 1.0pt,line width=0.09mm] (\AliceZeroY,0,\AliceZeroX) -- (\WillieZeroY,0,\WillieZeroX) node[anchor=south east] {};
          \fill[BobColor, opacity=0.5] 
          (\AliceZeroY-0.85,0,\AliceZeroX) -- (\AliceZeroY-0.58,0,\AliceZeroX) arc[start angle=180, end angle=109, radius=0.53] -- (0.225*\BobZeroY+0.775*\AliceZeroY,0,0.225*\BobZeroX+0.775*\AliceZeroX) -- (0.35*\BobZeroY+0.65*\AliceZeroY,0,0.35*\BobZeroX+0.65*\AliceZeroX) arc[start angle=115, end angle=167, radius=1.185];
          \node[] (thetaB) at (\AliceZeroY-0.85,+0.60pt,\AliceZeroX) {$\scriptstyle\theta_B$};
          \fill[WillieColor, opacity=0.5] (\AliceZeroY-0.58,0,\AliceZeroX) arc[start angle=180, end angle=73, radius=0.53] -- (0.16*\WillieZeroY+0.84*\AliceZeroY,0,0.16*\WillieZeroX+0.84*\AliceZeroX) -- (\AliceZeroY,0,\AliceZeroX);
          \node[] (thetaW) at (\AliceZeroY+0.1,0.7pt,\AliceZeroX) {$\scriptstyle\theta_W$};

        
        \end{tikzpicture}
        
    \caption{Reference scenario.}
    \label{Fig:CoverCommRIS}
\end{figure}

\subsection{Signal Transmission and Reception Model}
We consider the RIS aided covert communication system shown in Fig.~\ref{Fig:CoverCommRIS}, which consists of a legitimate transmitter (referred to as Alice) equipped with $N$ antennas, a legitimate receiver (referred to as
Bob) and a warden receiver (referred to as Willie) both equipped with a single-antenna.
The Alice antennas are arranged in a uniform planar array (UPA) with $N_H$ rows and $N_V$ columns, resulting in $N = N_HN_V$.
In this environment, a RIS is present with the goal of supporting the communication between Alice and Bob, preventing Willie from being aware about the communication between Alice and Bob, which is a typical covert communication scenario.
The RIS is equipped with $M$ passive reconfigurable elements, forming a UPA with $M_H$ rows and $M_V$ columns, where $M=M_HM_V$.
Hence, in the considered scenario, the objective of the RIS is to maximize the rate of the legitimate link 
with a constraint on Willie's ability to realize that a communication between Alice and Bob is taking place.

We denote by $\mathbf{t}_B \in \mathbb{C}^{1 \times M}$ and $\mathbf{t}_W \in \mathbb{C}^{1 \times M}$ the channel between the RIS and Bob and Willie, respectively, and by $\mathbf{S} \in \mathbb{C}^{M \times N}$ the channel between Alice and the RIS. Then, the direct channel between Alice and Willie and Alice and Bob are denoted by $\mathbf{h}_B \in \mathbb{C}^{1 \times N}$ and $\mathbf{h}_W \in \mathbb{C}^{1 \times N}$, respectively.
Hence, we can express the received signals by Bob and Willie, $\mathbf{y}_B$ and $\mathbf{y}_W$, as
\begin{IEEEeqnarray}{rCl}
    {y}_B & = & \sqrt{P_a}\mathbf{h}_B \mathbf{v}+\sqrt{P_a}\mathbf{t}_B \boldsymbol{\Delta}(\mathbf{z})\mathbf{S} \mathbf{v} \label{eq:received_signal_b} + n_B \\
    {y}_W & = &\sqrt{P_a}\mathbf{h}_W \mathbf{v} +\sqrt{P_a}\mathbf{t}_W \boldsymbol{\Delta}(\mathbf{z})\mathbf{S} \mathbf{v} + n_W \label{eq:received_signal_w}
\end{IEEEeqnarray}
where, $n_B \sim \mathcal{N}(0, \sigma^2_B)$ and $n_W \sim \mathcal{N}(0, \sigma^2_W)$ are the noise terms, $\boldsymbol{\Delta}(\mathbf{z}) \in \mathbb{C}^{M \times M}$ is the RIS reflecting matrix, dependent on a vector $\mathbf{z} \in \mathbb{C}^{M \times 1}$, $\mathbf{v} \in \mathbb{C}^{N \times 1}$ is the normalized precoding vector, i.e., $\mathbf{v}^H\mathbf{v} = 1$ and $P_a$ is the power transmitted by Alice.
Regarding the matrix $\mathbf{S}$, it is assumed to be perfectly known from the network deployment.
On the other hand, Alice is a legitimate user with a known position, and the RIS is typically in a fixed position.
Moreover, it is Alice's responsibility to position herself in order to establish a strong Line-of-Sight (LOS) connection with the RIS.
On the contrary, we make the reasonable assumption that the channel between the RIS and the malicious user Willie is not known on Alice's side.
To maintain generality, we also assume that Bob's channel is not perfectly known. 

\subsection{Channel model from Alice perspective}
In this section, we provide a characterization of the channel from Alice's perspective, specifically for the direct and reflected links toward Bob and Willie, whose states are not perfectly known to Alice.
In particular, based on the assumption that Alice knows the channel model and has an estimate of the positions of the nodes, we will derive the first and second order statistics of the channels of interest.
To achieve this, we will consider a classic statistical channel model valid in the far-field regime.
To maintain the generality of our approach, it is important to note that the optimization algorithm for the RIS and precoder relies on knowledge of the channel correlation matrices.
Therefore, even with a different channel model, such as in the near-field regime, the presented optimization framework remains applicable as long as Alice can compute these matrices using prior information or measurements.

In particular, given the potential presence of both Line of Sight (LOS) and non-Line of Sight (NLOS) components in practical context, we characterize all the involved channels using a generalized spatially correlated Rician fading model \cite{1499046}.
Spatially correlated Rician fading model is widely used in the literature to account for the presence of multipath fading in RIS-aided communications, as shown, for example, in \cite{RisScatteringParameterNetworkAnalysis} for a multiport network model using S-parameters.

In this setting, we consider a far-field propagation scenario where Alice knows that Willie and Bob are located in a certain area $\mathcal{P}_W$ and $\mathcal{P}_B$, respectively.
Therefore their positions, namely $\boldsymbol{\wp}_W$ and $\boldsymbol{\wp}_B$, follow some spatial distribution described by $f_{\boldsymbol{\wp}_W}(\mathbf{p})$ and $f_{\boldsymbol{\wp}_B}(\mathbf{p})$, respectively, known by Alice.

In the literature, channels involving Bob are typically assumed to be perfectly known since he is a 'friendly' user.
However, in RIS-aided communication systems, channel estimation is complex due to the passive nature of RIS elements and high-dimensional channels \cite{9328501, Demir2022is}.
Since RIS elements cannot transmit, receive, or process pilot signals, a central controller (e.g., the base station) must handle channel estimation, incurring a significant overhead compared to standard MIMO systems.
This challenge can be mitigated using statistical channel estimates, such as those based on Bob's position \cite{ABR, WuZhang2020, Abrardo_two}.
In this work, we consider the more general case where Alice has only statistical knowledge of the Bob's channel, with perfect CSI being a special case covered in Section \ref{sec:Probsolution}.

By conditioning Willie/Bob's position to a generic point generically denoted by $\mathbf{p}$, and indicating with $Q$ the generic entity (i.e., $Q=\{W,B\}$), Alice can determine the LOS angle of departure (AOD) $\theta_Q(\mathbf{p})$ of her  signals transmitted towards $Q$, corresponding to the direct link $\mathbf{h}_Q$.
Analogously, Alice can determine the LOS AOD $\vartheta_Q(\mathbf{p})$ for the signals transmitted from the RIS, corresponding to the reflected links $\mathbf{t}_Q$. The situation is illustrated Fig.~\ref{Fig:CoverCommRIS}.
Note that considering AODs as scalar quantities corresponds to the reasonable assumption that the distances between the nodes and between the nodes and the RIS are much greater than the heights, resulting in a planar geometry of the problem. 


To elaborate, let us focus on the reflected links, i.e., the channels between the RIS\footnote{This can be easily extended for the direct channels, where the signal comes from Alice, i.e., $\mathbf{h}_W$ and $\mathbf{h}_B$} and Wille/Bob, generically denoted as $\mathbf{t}_Q$.
Let us denote by $x_m$ the $x$ location of the $m$-th element of the RIS, with $m = 1,\ldots,M$, which are known by Alice.
When a plane-wave is reflected by the RIS, the array response vector toward the angle $\vartheta_Q(\mathbf{p})$ can be written as \cite{Demir2022Channel} 
\begin{equation}\label{array_vector_a}
    \mathbf{a}\big(\vartheta_Q(\mathbf{p})\big) = \left[e^{j\frac{2\pi}{\lambda}x_1 \sin\left(\vartheta_Q(\mathbf{p})\right) },\ldots,e^{j\frac{2\pi}{\lambda} x_M \sin \left(\vartheta_Q(\mathbf{p})\right)}\right]^T
\end{equation}
where $\lambda$ is the wavelength.

Then, the LOS component conditioned on $\mathbf{p}$, representing the conditional mean value of the Rice channel, is given by
\begin{equation}
    \boldsymbol{\mu}_{\mathbf{t}_Q}(\mathbf{p}) = \sqrt{\beta_{\mathbf{t}_Q}(\mathbf{p})\frac{K}{K+1}}\mathbf{a}\big(\vartheta_Q(\mathbf{p})\big)
\end{equation}
where $K$ is the Ricean factor, $\beta_{\mathbf{t}_Q}(\mathbf{p})$ encompasses the channel gain, evaluated assuming a line-of-sight (LOS) propagation model.
Both $K$ and $\beta_{\mathbf{t}_Q}(\mathbf{p})$ are supposed to be known by Alice.

Note that the model adopted for the LOS component is the same as the one considered in \cite{Abrardo_two}, in which it is shown that when the nodes' uncertainty is at least some wavelengths, the LOS component can be reasonably assumed to be a zero-mean Gaussian term. In this work, the assumption of Gaussianity is not necessary; instead, we maintain the assumption that the LOS components are zero-mean. The zero-mean assumption is less stringent than the Gaussian assumption and remains valid even for regions of uncertainty of a few wavelengths or fraction of  a wavelength. Accordingly, denoting the average of a generic matrix $\mathbf{M}$, which depends on $\mathbf{p}$, by
\begin{equation}
    \mathbb{E}_{\boldsymbol{\wp}_Q}\big[\mathbf{M}(\mathbf{p})\big] = \displaystyle\iint^{}_{\mathcal{P}_W} \mathbf{M}(\mathbf{p}) f_{\boldsymbol{\wp}_Q}(\mathbf{p})d\mathbf{p},
\end{equation}
we have $\mathbb{E}_{\boldsymbol{\wp}_Q}\left[\boldsymbol{\mu}_{\mathbf{t}_Q}(\mathbf{p})\right] = 0$. 

Regarding the NLOS modeling of the spatial correlation matrix, we adopt the approach proposed in \cite{Bjornson2021Rayleigh,Demir2022Channel} in which the NLOS is modeled as zero-mean Gaussian term given by a continuous superposition of paths transmitted with different AODs within a certain angular spread.
Specifically, considering a uniform angular distribution and denoting by $\Delta_{m}$ the angular spread, conditioning on the effective angular position of the node $\vartheta_Q$, the correlation matrix of the NLOS part is given by
\begin{equation}\label{Rhmp}
    \boldsymbol{\Sigma}_{\mathbf{t}_Q}\big(\vartheta_Q(\mathbf{p})\big) = \frac{\beta_\mathbf{\mathbf{t}_Q}(\mathbf{p})}{K+1}\!\int^{\vartheta_Q(\mathbf{p})+\Delta_{m}/2}_{\vartheta_Q(\mathbf{p})-\Delta_{m}/2} \mathbf{a}(\varphi)\mathbf{a}^{\text{H}}(\varphi)d\varphi.
\end{equation}
so from \eqref{array_vector_a} the generic element can be written as
\begin{equation}
    \left[\boldsymbol{\Sigma}_{\mathbf{t}_Q}\right]_{i,j}=\frac{\beta_\mathbf{\mathbf{t}_Q}(\mathbf{p})}{K+1}\!\int^{\vartheta_Q(\mathbf{p})+\Delta_{m}/2}_{\vartheta_Q(\mathbf{p})-\Delta_{m}/2} e^{j \frac{2\pi}{\lambda}((x_i - x_j) \sin (\varphi))} d\varphi.
\end{equation}

Assuming that Alice knows $\Delta_m$, it is possible to compute each element of $\boldsymbol{\Sigma}_{\mathbf{t}_Q} \big( \vartheta_Q(\mathbf{p}) \big)$ by using simple numerical integration.
Similar principles apply to the direct links, where the steering vector $\mathbf{a}(\cdot)$, defined in equation \eqref{array_vector_a}, is calculated with respect to the AODs from Alice antenna array rather than the RIS, i.e., using $\theta_W(\mathbf{p})$ and $\theta_B(\mathbf{p})$.

To sum up, we assume Alice has knowledge of:
\begin{itemize}
    \item Willie and Bob's spatial distribution, $f_{\wp_Q}(\mathbf{p})$;
    \item the direct channel with the RIS ($\mathbf{S}$), as well as the RIS geometry and position;
    \item the Ricean factor ($K$), loss model ($\beta$), and multipath angular spread ($\Delta_m$) of the channels (both direct and reflected) toward Willie and Bob;
    \item Willie’s noise model (see Sec.~\ref{sec:WillieDetection}).
\end{itemize}

Note that the assumptions made about channel knowledge are very general and cover several cases currently discussed in the literature.
As for the case where the channel to Bob is perfectly known, it simply involves replacing the channel correlation matrices with the products of the channels, as will be shown in Section \ref{sec:Probsolution}.
On the other hand, cases where only the mean and the variance of the path-losses involving Willie are known, as in \cite{CovertCommRIS,DelayContraintRISCC}, correspond to assuming $\Delta_m = 2\pi$ and $K = 0$.
In this regard, note that even though Willie is a malicious node for which the channel cannot be estimated, assuming certain statistics about his position is often a valid assumption, such as when Alice knows that Willie can only occupy a certain area, or if Alice is listening to Willie's transmissions and can determine his position or angle of arrival with a certain approximation.

The first and second-order statistics of the reflected channels can be computed as
\begin{IEEEeqnarray}{rCl}
    \boldsymbol{\mu}_{\mathbf{t}_Q} &=& \mathbb{E}_{\boldsymbol{\wp}Q}\left[\boldsymbol{\mu}_{\mathbf{t}_Q}(\mathbf{p})\right] = 0 \label{eq:mut}\\
    \mathbf{R}_{\mathbf{t}_Q} \nonumber &=& \mathbb{E}\left[\mathbf{t}_Q^H \mathbf{t}_Q\right]\\
    &=& \mathbb{E}_{\boldsymbol{\wp}_Q}\bigg[\frac{K}{1+K}\beta_{\mathbf{t}_Q}(\mathbf{p})\Big(\mathbf{a}\big(\vartheta_Q(\mathbf{p})\big)\mathbf{a}^{H}\big(\vartheta_Q(\mathbf{p})\big)\Big) + \nonumber\\
    &&\quad\quad \boldsymbol{\Sigma}_{\mathbf{t}_Q}\left(\vartheta_Q(\mathbf{p})\right) \bigg] \label{eq:RtQ}
\end{IEEEeqnarray}
and similarly for the direct channels we have
\begin{IEEEeqnarray}{rCl}
    \boldsymbol{\mu}_{\mathbf{h}_Q} &=& \mathbb{E}_{\boldsymbol{\wp}Q}\left[\boldsymbol{\mu}_{\mathbf{h}_Q}(\mathbf{p})\right] = 0 \label{eq:muh}\\
    \mathbf{R}_{\mathbf{h}_Q} \nonumber &=& \mathbb{E}\left[\mathbf{h}_Q^H \mathbf{h}_Q\right] \nonumber \\
    &=& \mathbb{E}_{\boldsymbol{\wp}_Q}\bigg[\frac{K}{1+K}\beta_{\mathbf{h}_Q}(\mathbf{p})\Big(\mathbf{a}\big(\theta_Q(\mathbf{p})\big)\mathbf{a}^{H}\big(\theta_Q(\mathbf{p})\big)\Big) + \nonumber\\
    &&\quad\quad \boldsymbol{\Sigma}_{\mathbf{h}_Q}\left(\theta_Q(\mathbf{p})\right) \bigg].\label{eq:RhQ}
\end{IEEEeqnarray}

Alice is finally able to calculate the normalized average powers received by Bob and Willie as
\begin{IEEEeqnarray}{rl}
    {P}_B(\mathbf{\Phi},\mathbf{v}) & = \mathbb{E}\left[\mathbf{v}^H \left(\mathbf{h}_B+\mathbf{t}_B\boldsymbol{\Phi} \right)^H\left(\mathbf{h}_B+\mathbf{t}_B\boldsymbol{\Phi} \right) \mathbf{v}\right]\label{Pot_avg1Bob}\\
    {P}_W(\mathbf{\Phi},\mathbf{v}) & = \mathbb{E}\left[\mathbf{v}^H \left(\mathbf{h}_W+\mathbf{t}_W\boldsymbol{\Phi} \right)^H\left(\mathbf{h}_W+\mathbf{t}_W\boldsymbol{\Phi} \right) \mathbf{v}\right].\label{Pot_avg1Willie}
\end{IEEEeqnarray}

Making the reasonable assumption that the direct and reflected path components from the RIS are uncorrelated, from \eqref{eq:mut}, \eqref{eq:RtQ}, \eqref{eq:muh}, \eqref{eq:RhQ}, we obtain
\begin{IEEEeqnarray}{rl}
    {P}_B(\mathbf{\Phi},\mathbf{v}) &= \mathbf{v}^H \left(\mathbf{R}_{{\mathbf{h}_B}}+\boldsymbol{\Phi}^H \mathbf{R}_{{\mathbf{t}_B}} \boldsymbol{\Phi}\right) \mathbf{v}\label{eq1bBob}\\
    {P}_W(\mathbf{\Phi},\mathbf{v}) &= \mathbf{v}^H \left(\mathbf{R}_{{\mathbf{h}_W}}+\boldsymbol{\Phi}^H \mathbf{R}_{{\mathbf{t}_W}} \boldsymbol{\Phi}\right) \mathbf{v}.\label{eq1bWillie}
\end{IEEEeqnarray}

In the specific case where Alice knows the channel to Bob exactly, the power $P_B$ becomes an instantaneous power.
In order to avoid abuse of notations, we use the same symbol and define it in this case as
\begin{equation}\label{Pot_CSI}
 {P}_B(\mathbf{\Phi},\mathbf{v}) = \mathbf{v}^H \left(\mathbf{h}_B+\mathbf{t}_B\boldsymbol{\Phi} \right)^H\left(\mathbf{h}_B+\mathbf{t}_B\boldsymbol{\Phi} \right) \mathbf{v}    
\end{equation}
\subsection{Willie's Detection on Alice's side}\label{sec:WillieDetection}
Assuming that Willie uses a radiometer detector, which is commonly used in practice to detect the presence of a signal \cite{CovertCommRIS,MIMORISCC,RandomWirelessNetworks}, the decision rule for Willie is based solely on the power level of the received signal.
Willie's detection is usually modelled as a binary hypotheses testing \cite{CovertCommRIS}, where $\mathcal{H}_0$ represents the null hypothesis, i.e., absence of signal from Alice, and the alternative hypothesis $\mathcal{H}_1$ indicates the presence of signal.
Here we focus on the worst-case assumption where Willie knows the channel states, the RIS coefficients $\boldsymbol{\Phi}$, Alice's precoder $\mathbf{v}$, and Alice's transmitting power $P_a$.
Conversely, as commonly assumed in covert communication study, noise power at Willie's receiver, $\sigma_W^2$, is not perfectly known and can be described by the bounded uncertainty model \cite{CovertCommNoiseUncertainty}, according to which the probability density function of the noise power at Willie's receiver is
\begin{equation}\label{eq:NoisePdf}
    f_{\sigma_W^2}(x)=\left\{
    \begin{array}{ll}
         \frac{1}{2x\ln(\rho)}& \rho^{-1}\tilde{\sigma}_W^2\leq x \leq \rho\tilde{\sigma}_W^2 \\
         0& \mbox{otherwise}
    \end{array}
    \right.
\end{equation}
where $\rho\geq1$ denotes the uncertainty coefficient and $\tilde{\sigma}_W^2$ is a reference noise power level.

The average power of the signal received by Willie with respect to the noise power, under the two hypotheses, is given by
\cite{CovertCommRIS}
\begin{equation}
    \Omega_W = \left\{
    \begin{array}{ll}
        \sigma_W^2 & \quad\mathcal{H}_0 \\
        P_a\mu+\sigma_W^2 & \quad\mathcal{H}_1
    \end{array}
    \right.\label{eq:OmegaW}
\end{equation}
with $\mu = \left|\big(\mathbf{h}_W +\mathbf{t}_W \boldsymbol{\Phi}\big) \mathbf{v}\right|^2$.
Note that $P_a\mu$ is the instantaneous power assumed to be known by Willie in the worst-case scenario, while $P_W$ in \eqref{eq1bWillie} is the average normalized power that can be computed on Alice's side, given her partial knowledge of the channel between Alice and Willie.

By denoting with $\Gamma$ the detection threshold used by Willie, assuming equal a-priori probabilities of $\mathcal{H}_0$ and $\mathcal{H}_1$, the detection error probability (DEP) is given by the sum of the false alarm and miss-detection probabilities, that is \cite{CovertCommRIS}
\begin{equation}\label{DEP_base}
    \zeta = 1-F_{\sigma_W^2}\left(\Gamma\right)+F_{\sigma_W^2}\left(\Gamma-P_a\mu\right)
\end{equation}
where $F_{\sigma_W^2}(\cdot)$ is the cumulative distribution function of $\sigma_W^2$.
{It is worth noting that in the literature, there are alternative expressions for DEP that arise from different detection strategies employed by Willie. An instance of this is considered in \cite{DelayContraintRISCC}, where Willie's decision can tolerate a given delay. In this case, the decision can be based on the KL divergence from $\mathbb{P}_0$ to $\mathbb{P}_1$, i.e., the likelihood functions of Willie's observation vector over independent channel uses under $\mathcal{H}_0$ and $\mathcal{H}_1$, respectively. In the following, we will refer to the baseline strategy presented in \eqref{DEP_base}, which captures the essence of the problem. However, as discussed in Section ~\ref{sec:ApproachGeneralization}, the proposed optimization strategy for Alice can also be extended to other scenarios.}

Accordingly, as shown in \cite{CovertCommRIS}, the optimal detection threshold that allows Willie to minimize the DEP turns out to be
\begin{equation}
    \Gamma^*(P_a\mu) = \min\left(P_a\mu+\tilde{\sigma}_W^2\rho^{-1},\tilde{\sigma}_W^2\rho\right),
\end{equation}
and the related minimum DEP is
\begin{equation}
    \zeta^*(P_a\mu) = \left\{
    \begin{array}{cc}
        1-\frac{1}{2}\log_\rho \left(1+\frac{\rho P_a\mu}{\tilde{\sigma}_W^2}\right) &  \mu \leq \frac{\tilde{\sigma}_W^2}{P_a}\left(\rho-\rho^{-1}\right)\\
        0 & \mu > \frac{\tilde{\sigma}_W^2}{P_a}\left(\rho-\rho^{-1}\right)
    \end{array}
    \right.\label{eq:minDEP}
\end{equation}

Looking at \eqref{eq:minDEP}, it is clear that DEP higher than zero is achievable only if $\rho\frac{P_a\mu}{\tilde{\sigma}_W^2}<\rho^2-1$.
In practical scenarios where the DEP approaches $1$, equation \eqref{eq:minDEP} can be simplified.
For instance, from \eqref{eq:minDEP} in order to have $\zeta^*(P_a\mu) > 1-\delta$ with $\delta << 1$, we have
\begin{equation}
    \rho\frac{P_a\mu}{\tilde{\sigma}_W^2}<\rho^{2\delta}-1\approx 2\delta << 1,
\end{equation}
which yields to
\begin{equation}
    \zeta^*(P_a\mu) \approx 1-\frac{1}{2\log(\rho)}\frac{\rho P_a\mu}{\tilde{\sigma}_W^2}\label{eq:ZetaApprox}.
\end{equation}

Note that from the relationship $\log(1+x) \le x$, the expression for the DEP that derives from the approximation \eqref{eq:ZetaApprox} always provides a lower bound on the DEP, meaning that it represents a more conservative sub-optimal approach.
Therefore, equation \eqref{eq:ZetaApprox} can be considered by Alice to establish the covertness constraint.
However, the minimum DEP is not available on Alice's side, as only partial information of the channels is available.
On the other hand, Alice can exploit the information about the channel statistic, and therefore compute the average DEP, which is given by
\begin{IEEEeqnarray}{rCl}
    \overline{\zeta}(P_a,\mathbf{v},\boldsymbol{\Phi}) & = & \mathbb{E}_{\mathbf{h}_W,\mathbf{t}_W,\boldsymbol{\wp}_W}\left[\zeta^*\right] \nonumber \label{eq:avgZeta}\\
    & \approx & 1-\frac{1}{\log(\rho)}\frac{\rho P_a P_W(\boldsymbol{\Phi},\mathbf{v})}{\tilde{\sigma}_W^2}.\label{eq:avgZetaInt}
\end{IEEEeqnarray}

Therefore, the average covert constraint, i.e., $\overline{\zeta}<\zeta_{min} = 1-\delta_{min}$, can be expressed in terms of average power received by Willie as
\begin{equation}
    P_aP_W(\boldsymbol{\Phi},\mathbf{v})<\frac{\delta_{min}}{\log(\rho)\frac{\tilde{\sigma}_W^2}{\rho}}=Q_{max}.\label{eq:PowerCovertConstr}
\end{equation}

Note that the approach of considering the average DEP as a constraint is usually considered in works where Alice has only partial knowledge of the channel, for example, see \cite{CovertCommRIS,DelayContraintRISCC,ActivePassiveBeamFormingRISCC}

\section{RIS Model}\label{Models}
In this Section, we provide a general model of a RIS that highlights the shortcomings of traditionally used models. To achieve this, we start from a classic model to represent a RIS within a communication system. Specifically, it is customary to represent a RIS as a surface composed of ideal scatterers capable of imparting a phase shift to the impinging signal. This model is referred to as the Communication Theory (CT) model and is depicted below. It is then shown that this model is a special case of a more general and accurate model that we refer to as the multiport (MP) network model. The comparison between the two models is subsequently discussed, highlighting the limitations of the CT model.

\subsection{Communication Theory (CT) Model}
In the CT model, the RIS is represented by a diagonal reflection matrix with elements:
\begin{equation}\label{eq:TFZ_CT2}
\left[\boldsymbol{\Delta}_{CT}\right]_{m,m} = J_0 \Gamma_m
\end{equation}
where $J_0$ is a constant, and $\Gamma_m = e^{j\phi_m}$ is the load reflection coefficient with $0 \le \phi_m \le 2\pi$. 

\subsection{MP Model}
By referring to the MP $Z$-parameters representation of an end-to-end channel matrix involving an RIS as presented in \cite{abrardo_E}, we obtain:
\begin{equation}\label{eq:TFZ}
\boldsymbol{\Delta}_{MP}(\mathbf{z}) = -2Y_0 (\mathbf{Z}_{SS}+r_0 \mathbf{I}_M+\mathbf{Z}_R)^{-1}.
\end{equation}
Here, $\mathbf{Z}_{SS} \in \mathbb{C}^{M \times M}$ denotes the matrix of self and mutual impedances of the RIS, $\mathbf{Z}_R = \text{diag}(\mathbf{z})$, where $\mathbf{z} = j\mathbf{b}$ are tunable impedances connected at the RIS ports, $r_0$ represents a small parasitic resistance, $\mathbf{I}_M$ stands for the identity matrix, and $Y_0$ is a reference admittance defined as:
\begin{equation}\label{eq:Y0}
Y_0 = \frac{Z_0}{(Z_0+Z_{RR})(Z_0+Z_{TT})}.
\end{equation}
In \eqref{eq:Y0}, $Z_0$ is the reference impedance across which the transmitter and receiver ports are terminated, while $Z_{RR}$ and $Z_{TT}$ denote the self-impedances of the receiver and transmitter ports. For simplicity, it is assumed that $Z_0$ is uniform across all ports and equals to $Z_0 = 50$ $\Omega$. Note that \eqref{eq:TFZ} has dimensions of $\Omega^{-2}$, consistent with the Z-parameter representation of the MP model, where $\mathbf{S}$ and $\mathbf{t}_i$ in \eqref{eq:received_signal_w} denote impedance matrices with dimensions of $\Omega$. The values of $\mathbf{Z}_{SS}$, $Z_{RR}$, and $Z_{TT}$ can be analytically computed as demonstrated in \cite{GradoniEnd2End}, or determined using full-wave simulators as shown in \cite{abrardo_E}; nevertheless, they can be assumed to be known since they rely on the structural characteristics of the RIS, such as the length and spacing between dipoles (in the case of RIS composed of dipoles). 

\subsection{Comparison between CT and MP models}
The CT model in \eqref{eq:TFZ_CT2} can be seen as a particular case of the more general MP model in \eqref{eq:TFZ}. It is indeed straightforward to derive the CT model from the MP one by making the following assumptions.
First we neglect the parasitic resistance $r_0$ and impose $\mathbf{Z}_{SS} = Z_0 \mathbf{I}_M$.
This simplified model is denoted by ideal Multi-Port (iMP) which, from \eqref{eq:TFZ}, yields a diagonal RIS reflection matrix with elements:
\begin{equation}\label{eq:TFZ_CT1}
    \left[\boldsymbol{\Delta}_{iMP}(\mathbf{b})\right]_{m,m} = -2Y_0 ({Z}_{0}+jb_m)^{-1}. 
\end{equation}
Hence, it is easy to verify that
\begin{equation}\label{eq:TFZ_CT1bis}
    \boldsymbol{\Delta}_{CT}(\mathbf{b}) = \boldsymbol{\Delta}_{iMP}(\mathbf{b}) + \frac{Y_0}{{Z}_{0}} \mathbf{I}_M,
\end{equation}
when $J_0 = \frac{Y_0}{Z_0}$ and 
\begin{equation}\label{eqGamma}
    \Gamma_m = \frac{jb_m-Z_0}{jb_m+Z_0} \rightarrow \phi_m = - 2 \tan^{-1} \frac{b_m}{Z_0} + \pi
\end{equation}
in \eqref{eq:TFZ_CT2}.
The above equation sheds light on a third important approximation that is inherent in the CT model, in which the term $\Delta_{\mathbf{S}} = \frac{Y_0}{Z_0} \mathbf{I}_M$ is added to the iMP model. To understand the physical significance of this discrepancy, consider the iMP model in \eqref{eq:TFZ_CT1} and assume that the RIS is connected to a resistance $z_m = Z_0$. In this case, the response of the RIS according to the iMP model is given by $-\Delta_{\mathbf{S}}$, while for the CT model, the response is zero. This component, which is always present in reality, represents a structural component that causes the RIS to behave like a scatterer that does not introduce any phase shift on the incident signal, thereby introducing a specular component that, as we will see, cannot be ignored without significantly deteriorating performance in certain situations.

Furthermore, the CT model ignores the possible mutual coupling between the elements of the RIS and the fact that it is impossible to control the phase without inevitably varying the amplitudes of the responses of the individual RIS elements, as highlighted by the MP model in \eqref{eq:TFZ}.
In this regard, in some works in the literature, instead of assuming fixed amplitudes as in the CT model, the amplitudes of the RIS element responses are assumed to be controllable independently of the phases (e.g., see \cite{DelayContraintRISCC}). The model in \eqref{eq:TFZ} shows that, although there is some flexibility in varying the amplitudes of the RIS element responses, it is not possible to have independent control over both amplitude and phase.

From now on, we will consider the general MP model in \eqref{eq:TFZ} which also includes the CT model. 
For simplicity, we denote $\boldsymbol{\Delta}(\mathbf{b}) = \boldsymbol{\Delta}(j\mathbf{b})$, where $\mathbf{b} \in \mathbb{R}^M$ represents the tunable reactances connected to the RIS ports. We also use $\boldsymbol{\Phi} = \boldsymbol{\Delta}(\mathbf{b}) \mathbf{S} \in \mathbb{C}^{M \times N}$ to simplify the notation where possible.

\section{Problem Formulation}\label{sec:ProblemFormulation}

The conventional covert communication problem (CCP) aims to maximize Bob's rate with a cover constraint toward Willie. As for Bob's rate, considering the availability of the statistical CSI, we can define the ergodic rate as
\begin{equation}\label{ERG_Rate}
    R = \mathbb{E}_{\mathbf{h}_B,\mathbf{t}_B,\boldsymbol{\wp}_B} \log_2\left\{1+\gamma(\mathbf{h}_B,\mathbf{t}_B,\mathbf{p},\boldsymbol{\Phi},\mathbf{v})\right\}
\end{equation}
where
\begin{IEEEeqnarray}{rl}
    \gamma(\mathbf{h}_B,&\mathbf{t}_B,\mathbf{p},\boldsymbol{\Phi},\mathbf{v}) = \label{ERG_Gamma}\\
    & \frac{P_a\mathbf{v}^H \left(\mathbf{h}_B(\mathbf{p})+ \mathbf{t}_B(\mathbf{p})\boldsymbol{\Phi}\right)\left(\mathbf{h}_B(\mathbf{p})+ \mathbf{t}_B(\mathbf{p})\boldsymbol{\Phi}\right)^H \mathbf{v}}{\sigma^2_B}.\nonumber
\end{IEEEeqnarray}

We consider the upper bound $R_b \geq R$ derived from the concavity of the logarithm function, where
\begin{IEEEeqnarray}{rCl}\label{ERG_Rate1}
    R_b &=& \text{log}_2\left\{1+\mathbb{E}_{\mathbf{h}_B,\mathbf{t}_B,\boldsymbol{\wp}_B}\left[\gamma(\mathbf{h}_B,\mathbf{t}_B,\mathbf{p},\boldsymbol{\Phi},\mathbf{v})\right]\right\} \nonumber \\
    &=& \text{log}_2\left\{1+\frac{P_a{P}_B(\mathbf{\Phi},\mathbf{v})}{\sigma^2_B}\right\}
\end{IEEEeqnarray}
where ${P}_B(\mathbf{\Phi},\mathbf{v})$ is the average Bob's power given in \eqref{eq1bBob}. In the case where the channel to Bob is known, ${P}_B(\mathbf{\Phi},\mathbf{v})$ is the instantaneous power defined in \eqref{Pot_CSI}, and in this case, maximizing the power corresponds to maximizing the instantaneous rate.
Therefore, considering the availability of partial channel information and the equivalent covert constraint given by \eqref{eq:PowerCovertConstr}, and noting that the bound $R_b \ge R$ is very tight in practical scenarios where Bob's power is affected by low variance, we have:
\begin{IEEEeqnarray}{rCl}
    \hspace{-1.27cm}(\mbox{P0}):\quad \left(\mathbf{\Phi}^*,\mathbf{v}^*,P_a^*\right) &=& \arg \max_{\mathbf{\Phi},\mathbf{v},P_a}\ P_a{{P}_B(\mathbf{\Phi},\mathbf{v})}\label{eq:P0:CovertCommOptProb}\\
    \IEEEyessubnumber \textrm{s.t.:}&& \mathbf{\Phi}\in\mathcal{F}\\
    \IEEEyessubnumber && P_a  \mathbf{v}^H\mathbf{v} \le P_{max}\label{eq:Pconstraint} \\
    \IEEEyessubnumber && \overline{\zeta}\geq\zeta_{min}\label{eq:CovertConstraint}
\end{IEEEeqnarray}
where $P_{max}$ is the maximum transmittable power, \eqref{eq:CovertConstraint} represents the covertness constraint equivalent to ${P_a}{P}_W(\mathbf{\Phi},\mathbf{v}) \le {Q_{max}}$, according to \eqref{eq:PowerCovertConstr}, and the feasible set 
\begin{equation}
    \mathcal{F}=\{\mathbf{\Phi} \in \mathbb{C}^{M \times N}\vert \mathbf{\Phi} = \mathbf{S} \boldsymbol{\Delta}(\mathbf{b})\}
\end{equation}
enforces the multiport network model, where $\mathbf{b}$ represents the tunable reactances of the RIS, i.e., $\mathbf{b} \in \mathbb{R}^M$.


Hence, let us consider the following modified problem
\begin{IEEEeqnarray}{rCl}
    \hspace{-1cm}(\mbox{P1}):\quad \left(\hat{\mathbf{\Phi}}^*,\hat{\mathbf{v}}^*,\hat{P_a}^*\right) &=& \arg \max_{\mathbf{\Phi},\mathbf{v},P_a}\ P_a{{P}_B(\mathbf{\Phi},\mathbf{v})}\label{eq:P1:CovertCommOptProbNoPowConstr}\\
    \IEEEyessubnumber \textrm{s.t.:}&&\mathbf{\Phi}\in\mathcal{F}\\
    \IEEEyessubnumber && \mathbf{v}^H\mathbf{v} \le 1 \\
    \IEEEyessubnumber && {P_a}{P}_W(\mathbf{\Phi},\mathbf{v}) = {Q_{max}},\label{eq:PowerCovertConstrP1}
\end{IEEEeqnarray}
where the covertness constraint is modified considering the approximation \eqref{eq:PowerCovertConstr} and it is set to equality since it is easy to see that if we take any $P_a$ less than the one that meets the constraint at equality, it cannot be optimal because it would yield a lower rate for Bob.

Note that in (P1) we have neglected the constraint \eqref{eq:Pconstraint} in (P0), assuming that the limit on the maximum transmittable power is given only by the constraint on $Q_{max}$.
This corresponds to the scenario of practical interest, where Willie's presence forces Alice to limit the transmitted power compared to its maximum.

Here we prove the equivalence of problem (P1) in \eqref{eq:P1:CovertCommOptProbNoPowConstr} with the following  problem:
\begin{IEEEeqnarray}{rCl}
    \hspace{-2.1cm}(\mbox{P2}):\quad \left(\tilde{\mathbf{\Phi}}^*,\tilde{\mathbf{v}}^*\right) &=& \arg \max_{\mathbf{\Phi},\mathbf{v}}\ \frac{{P}_B(\mathbf{\Phi},\mathbf{v})}{{P}_W(\mathbf{\Phi},\mathbf{v})},\label{eq:P2:RatioOptProb}\\
    \IEEEyessubnumber \textrm{s.t.:}&&\mathbf{\Phi}\in\mathcal{F}\\
    \IEEEyessubnumber && \mathbf{v}^H\mathbf{v} \le 1,
\end{IEEEeqnarray}

The equivalence between (P1) and (P2) is proven through the following proposition:

\begin{prop}\label{theo2}
The triplet $\left(\tilde{\mathbf{\Phi}}^*,\tilde{\mathbf{v}}^*,\tilde{P}^*_a\right)$ is an optimal solution of (P1) when $\tilde{P}^*_a = \frac{Q_{max}}{{P}_W(\tilde{\mathbf{\Phi}}^*,\tilde{\mathbf{v}}^*)}$.
\end{prop}
\begin{IEEEproof}
Firstly, it is evident that if $P_a = \tilde{P}^*_a$, the constraint on $Q_{max}$ is satisfied.
Hence, since $\left(\hat{\mathbf{\Phi}}^*,\hat{\mathbf{v}}^*,\hat{P_a}^*\right)$ is an optimal solution for (P1), we have
\begin{equation}\label{opt_eq11_1}
    \tilde{P}_a^*{P}_B(\tilde{\mathbf{\Phi}}^*,\tilde{\mathbf{v}}^*) \le \hat{{P}}_a^*{P}_B({\hat{\mathbf{\Phi}}}^*,{\hat{\mathbf{v}}}^*).
\end{equation}
On the other hand, since $\left(\tilde{\mathbf{\Phi}}^*,\tilde{\mathbf{v}}^*\right)$ is an optimal solution for (P2), we have
\begin{equation}\label{opt_eq1}
    \frac{\tilde{P}_a^*{P}_B(\tilde{\mathbf{\Phi}}^*,\tilde{\mathbf{v}}^*)}{\tilde{P}_a^*{P}_W(\tilde{\mathbf{\Phi}}^*,\tilde{\mathbf{v}}^*)} \ge \frac{\hat{P}_a^*{P}_B({\hat{\mathbf{\Phi}}}^*,{\hat{\mathbf{v}}}^*)}{\hat{{P}}_a^*{P}_W({\hat{\mathbf{\Phi}}}^*,{\hat{\mathbf{v}}}^*)}.
\end{equation}
Since the denominators of \eqref{opt_eq1} are both equal to $Q_{max}$, we have
\begin{equation}\label{opt_eq11_2}
    \tilde{P}_a^*{P}_B(\tilde{\mathbf{\Phi}}^*,\tilde{\mathbf{v}}^*) \ge \hat{{P}}_a^*{P}_B({\hat{\mathbf{\Phi}}}^*,{\hat{\mathbf{v}}}^*).
\end{equation}
Accordingly, from \eqref{opt_eq11_1} and \eqref{opt_eq11_2},
we have $\tilde{P}_a^*{P}_B(\tilde{\mathbf{\Phi}}^*,\tilde{\mathbf{v}}^*) = \hat{{P}}_a^*{P}_B({\hat{\mathbf{\Phi}}}^*,{\hat{\mathbf{v}}}^*)$, which concludes the proof.
 \end{IEEEproof}
Hence, in the following we will consider problem (P2), where there is no explicit constraint on covertness, which can be achieved optimally by imposing ${P}_a = \frac{Q_{max}}{{P}_W(\tilde{\mathbf{\Phi}}^*,\tilde{\mathbf{v}}^*)}$.

Finally, if the resulting power $P_a$ exceeds $P_{\text{max}}$, $P_{\text{max}}$ can be used as the transmitted power, as it allows for satisfying the covertness constraint in this case.
In this scenario, the solution to problem (P2) may be sub-optimal compared to case (P0).

\subsection{Approach generalization}\label{sec:ApproachGeneralization}
As highlighted in Section~\ref{sec:WillieDetection}, Willie could implement more elaborate strategies compared to the one that leads to DEP in \eqref{DEP_base}, making the calculation of DEP a complex function of Alice's transmission parameters $\mathbf{\Phi}$, $\mathbf{v}$, and $P_a$.
Nevertheless, the approach proposed in this work, based on problem separation, remains valid, although it leads to a sub-optimal solution. This solution essentially corresponds to making the assumption that the average DEP is a linearly decreasing function of the average power received by Willie for calculating $\mathbf{\Phi}$ and $\mathbf{v}$. This allows finding $P_W$ by solving problem (P2). Thus, the calculation of $P_a$ can be obtained in a second stage by considering the actual DEP model provided by Willie's strategy.

\section{Problem solution}\label{sec:Probsolution}
Problem (P2) in \eqref{eq:P2:RatioOptProb} is more tractable but still not convex.
Therefore, we present an alternative formulation of the same problem below, enabling us to find a local optimum.
The proposed approach is based on alternating optimization (AO).
The AO algorithm, known for breaking down optimization tasks into smaller, more manageable subproblems, simplifies the optimization process.
This approach proves particularly useful when dealing with intricate interactions or dependencies among variables.
Specifically, a three-steps algorithm is proposed in which, in addition to the two variables $\mathbf{v}$ and $\mathbf{\Phi}$ present in \eqref{eq:P2:RatioOptProb}, there is an auxiliary variable $\boldsymbol{\lambda}$ that will be introduced later.
These variables are alternately optimized, i.e., one of the three variables is optimized assuming the other two fixed.
To start, let's consider being at iteration $k$ of the AO algorithm, and we denote by $\mathbf{v}^{(k)}$ and $\mathbf{\Phi}^{(k)}$ the values of $\mathbf{v}$ and $\mathbf{\Phi}$ found at iteration $k$.
Then, a method is derived for computing $\mathbf{\Phi}^{(k+1)}$ with the precoder fixed to $\mathbf{v}^{(k)}$, which for simplicity of notation is simply denoted as $\mathbf{v}$.

\subsection{RIS optimization}\label{sec:RisOptimization}
In the following, we describe a general approach for RIS optimization that is valid for both the MP model and the CT model.
To elaborate, given the precoding vector $\mathbf{v}$, we formulate problem (P2) in \eqref{eq:P2:RatioOptProb} as follows
\begin{IEEEeqnarray}{rCl}
    \mathbf{\Phi}^{(k+1)} &=& \arg \max_{\mathbf{\Phi}}\ \frac{{P}_B(\mathbf{\Phi},\mathbf{v})}{{P}_W(\mathbf{\Phi},\mathbf{v})},\label{opt_2_3}\\
    \IEEEyessubnumber \textrm{s.t.:} && \mathbf{\Phi}\in\mathcal{F}
\end{IEEEeqnarray}

The problem \eqref{opt_2_3} is an example of fractional programming (FP), which is a non-convex type of problem \cite{FractionalProgramming}.
In the following, steps are developed to derive a manageable equivalent representation of the problem.
Let us consider the general case where the channel to Bob is not known deterministically.
We can observe that $\mathbf{R}_{\mathbf{h}_B}$ and $ \mathbf{R}_{\mathbf{t}_B}$ are semidefinite positive Hermitian matrices and as such they can be diagonalized as
\begin{IEEEeqnarray}{rCl}
    \mathbf{R}_{\mathbf{h}_B} &=& \mathbf{U}_{\mathbf{h}_B}\mathbf{D}_{\mathbf{h}_B} \mathbf{U}^H_{\mathbf{h}_B}\label{eq:SolOpt_2_3Bob}\\
    \mathbf{R}_{\mathbf{t}_B} &=& \mathbf{U}_{\mathbf{t}_B}\mathbf{D}_{\mathbf{t}_B} \mathbf{U}^H_{\mathbf{t}_B}\label{eq:SolOpt_2_3Willie}
\end{IEEEeqnarray}
where $
\mathbf{D}_{\mathbf{h}_B}$ and $ \mathbf{D}_{\mathbf{t}_B}$ have real non negative entries. Then, denote by $\mathbf{T}_{\mathbf{h}_B} = \mathbf{D}_{\mathbf{h}_B}^{\frac{1}{2}} \mathbf{U}^H_{\mathbf{h}_B} $ and $\mathbf{T}_{\mathbf{t}_B} = \mathbf{D}_{\mathbf{t}_B}^{\frac{1}{2}} \mathbf{U}^H_{\mathbf{t}_B} $ and introduce the matrix $\mathbf{Y}_{B} = \left[\mathbf{T}^T_{\mathbf{h}_B} ,\left(\mathbf{T}_{\mathbf{t}_B}\mathbf{\Phi}\right)^T\right]^T \in \mathbb{C}^{(M+N) \times N}$ obtained by stacking $\mathbf{T}_{\mathbf{h}_B} \in \mathbb{C}^{N \times N}$ and $\mathbf{T}_{\mathbf{t}_B}\mathbf{\Phi} \in \mathbb{C}^{M \times N}$ into a single matrix. From the above, it easily follows that $\mathbf{Y}^H_{B} \mathbf{Y}_{B} = \mathbf{R}_{\mathbf{h}_B}+ \mathbf{\Phi}^H\mathbf{R}_{\mathbf{t}_B}\mathbf{\Phi}$.
The above decomposition allows the use of the quadratic transform \cite{FractionalProgramming} to rewrite the problem \eqref{opt_2_3} as
\begin{IEEEeqnarray}{rCl}
    \mathbf{b}^{(k+1)} &=& \arg \min_{\mathbf{b}} \left[\min_{\boldsymbol{\lambda}} - \mathcal{G}\left(\boldsymbol{\lambda},\mathbf{b}, \mathbf{v} \right)\right],\label{opt_2_4}\\
    \IEEEyessubnumber \textrm{s.t.:} && \mathbf{b} \in \mathbb{R}^M
\end{IEEEeqnarray}
where $\boldsymbol{\lambda}\in\mathbb{C}^{1 \times M+N}$ is an auxiliary variable and
\begin{equation}\label{AUfun}
    \mathcal{G}\left(\boldsymbol{\lambda}, \mathbf{b}, \mathbf{v} \right) = 2\Re \left(\boldsymbol{\lambda} \mathbf{Y}_{B} \mathbf{v}\right) - \boldsymbol{\lambda}  {P}_W(\mathbf{\Phi}(\mathbf{b}),\mathbf{v}) \boldsymbol{\lambda}^H
\end{equation}

Indeed, the minimum of the inner problem is obtained when
\begin{equation}
    \boldsymbol{\lambda}^{*}(\mathbf{b},\mathbf{v}) = \arg \min_{\boldsymbol{\lambda}}\ -\mathcal{G}\left(\boldsymbol{\lambda},\mathbf{b}, \mathbf{v}  \right) = \frac{\mathbf{v}^H \mathbf{Y}^H_{B}}{ {P}_W(\mathbf{\Phi}(\mathbf{b}),\mathbf{v})}. \label{opt_2_10}
\end{equation}
Since $\mathbf{v}^H \mathbf{Y}^H_{B}\mathbf{Y}_{B}\mathbf{v} = {P}_B$, we finally have
\begin{equation}
\mathcal{G}\left(\boldsymbol{\lambda}^{*}(\mathbf{b},\mathbf{v}), \mathbf{b}, \mathbf{v} \right)  =  \frac{{P}_B(\mathbf{\Phi}(\mathbf{b}),\mathbf{v})}{{P}_W(\mathbf{\Phi(\mathbf{b})},\mathbf{v})}.\label{opt_2_1mm}
\end{equation}

As for the case where the instantaneous CSI to Bob is available, one can always write \eqref{AUfun}, \eqref{opt_2_10}, and \eqref{opt_2_1mm} simply by considering $\mathbf{T}_{\mathbf{h}_B} = \mathbf{h}_B$ and $\mathbf{T}_{\mathbf{t}_B} = \mathbf{t}_B$.\\ 
To solve this problem, following the AO approach, the two minimization problems are computed separately. Specifically, we proceed as follows:
\begin{enumerate}
\item With $\mathbf{b}=\mathbf{b}^{(k)}$, we minimize $-\mathcal{G}\left(\boldsymbol{\lambda},\mathbf{b}, \mathbf{v} \right)$ by setting $\boldsymbol{\lambda}^{(k+1)} = \boldsymbol{\lambda}^{*}(\mathbf{b}^{(k)},\mathbf{v}^{(k)})$ (see \eqref{opt_2_10}).
\item The RIS tunable impedance vector $\mathbf{b}^{(k+1)}$ is computed as the solution of the minimization problem: 
%
\begin{IEEEeqnarray}{rCl}
    \mathbf{b}^{(k+1)} &=& \arg \min_{\mathbf{b}}\ -\mathcal{G}\left(\boldsymbol{\lambda}^{(k+1)},\mathbf{b}, \mathbf{v} \right),\label{opt_2_6}\\
    \IEEEyessubnumber \textrm{s.t.:} && \mathbf{b} \in \mathbb{R}^M.
\end{IEEEeqnarray}
\end{enumerate}


The primary challenge presented by \eqref{opt_2_6} stems from the nonlinearity of $\mathbf{\Phi}(\mathbf{b}) = \mathbf{S}\Delta(\mathbf{b})$, where $\Delta(\mathbf{b})$ is defined in \eqref{eq:TFZ}.
To address this, rather than directly manipulating the tunable reactances, we opt to treat the phases of the reflection coefficients as tunable parameters.
To address the problem, we begin by assuming that the parasitic resistance $r_0$ is negligible, allowing us to express $\Gamma_m = \frac{jb_m - Z_0}{jb_m + Z_0}$, or $\Gamma_m = e^{j\phi_m}$, where $b_m = \frac{Z_0}{j} \frac{1 + e^{j\phi_m}}{1 - e^{j\phi_m}}$.
The proposed optimization approach for the RIS leverages the Neumann series approximation to linearize the matrix inverses for small variations of the solution.

Let $\mathbf{B}^{(k)} = \text{diag}\left(\mathbf{b}^{(k)}\right)$ and $\boldsymbol{\phi}^{(k)}$ denote the matrices of tunable reactances and the corresponding reflection coefficient phase vector at iteration $k$ of the iterative algorithm.
When assessing the solution at the next step, denoted as $\boldsymbol{\phi}^{(k+1)} = \boldsymbol{\phi}^{(k)} + \boldsymbol{\delta}$, where $\boldsymbol{\delta} \in \mathbb{C}^{1 \times M}$ and $\delta_m << 1$ to account for small variations in $\boldsymbol{\phi}^{(k)}$, we have
\begin{equation}\label{Bvar}
    \mathbf{B}^{(k+1)} \approx \mathbf{B}^{(k)} +\mathbf{F}^{(k)}\mbox{diag}\left(\boldsymbol{\delta}\right)
\end{equation}
where $\mathbf{F}^{(k)} \in \mathbb{C}^{M \times M}$ is a diagonal matrix with entries $\mathbf{F}^{(k)}_{m,m} =  \frac{d b_m^{(k)}}{d \phi_m} = -\frac{\left(b_m^{(k)}\right)^2+Z_0^2}{2Z_0}$.

To elaborate, for the sake of notation convenience, let us introduce $Q_1$, $Q_2$, $\mathbf{q}_1 \in \mathbb{C}^{N \times 1}$ and $\mathbf{q}_2 \in \mathbb{C}^{M \times 1}$ as
\begin{equation}\label{notations}
    \begin{array}{l}
        Q_1 = \mathbf{v}^H   \mathbf{R}_{{\mathbf{h}_W}}  \mathbf{v} \\
        Q_2 = \mathbf{v}^H \boldsymbol{\Phi}^H(\mathbf{b}) \mathbf{R}_{{\mathbf{t}_W}} \boldsymbol{\Phi}(\mathbf{b})  \mathbf{v} \\
        \mathbf{q}_1 = \mathbf{T}_{\mathbf{h}_B}\mathbf{v} \\
        \mathbf{q}_2 = \mathbf{T}_{\mathbf{t}_B} \mathbf{\Phi}(\mathbf{b}) \mathbf{v}
    \end{array}
\end{equation}
and denote $\boldsymbol{\lambda}^{(k+1)}_1 = {\lambda}^{(k+1)}_j$, with $j = 1,\ldots,N$ and $\boldsymbol{\lambda}^{(k+1)}_2 = {\lambda}^{(k+1)}_j$, with $j = N+1,\ldots,N+M$,  so that from \eqref{notations}, \eqref{AUfun} can be written as
\begin{IEEEeqnarray}{rCl}
    \mathcal{G}\left(\boldsymbol{\lambda}^{(k+1)},\mathbf{b}, \mathbf{v} \right) &=& \!\!2\Re \left( \boldsymbol{\lambda}^{(k+1)}_1 \mathbf{q}_1\right) + \!2\Re \left(\boldsymbol{\lambda}^{(k+1)}_2 \mathbf{q}_2\right) \nonumber\\
    && -\left \| \boldsymbol{\lambda}^{(k+1)} \right \|^2 Q_1 - \left \| \boldsymbol{\lambda}^{(k+1)} \right \|^2 Q_2 \nonumber\\\label{MSE_2}
\end{IEEEeqnarray}
Next, we introduce the following definitions:
\begin{IEEEeqnarray}{rCl}
    \mathbf{A}^{(k)} &=& \left(\mathbf{Z}_{SS}+r_0\mathbf{I}_M+j\mathbf{B}^{(k)}\right)^{-1} \label{notations3A}\\
    \mathbf{G}^{(k)} &=& j \mathbf{A}^{(k)}  \mathbf{F}^{(k)}.\label{notations3G}
\end{IEEEeqnarray}
From \eqref{Bvar}, by employing the Neumann series approximation for matrix inversion, we get
\begin{equation}\label{refeq2}
    (\mathbf{Z}_{SS}+r_0\mathbf{I}_M+j\mathbf{B}^{(k+1)})^{-1} \approx {\mathbf{A}}^{(k)} - \mathbf{G}^{(k)}\text{diag}\left(\boldsymbol{\delta}\right){\mathbf{A}}^{(k)}
\end{equation}
valid under the condition
\begin{equation}\label{VNcond}
    \left \| \mathbf{F}^{(k)}\text{diag}\left(\boldsymbol{\delta}\right) \mathbf{A}^{(k)} \right \| \le \epsilon
\end{equation}
where $\epsilon << 1$.
Denoting by $\mathbf{L} = \mathbf{F}^{(k)} \mathbf{A}^{(k)}$, condition \eqref{VNcond} can be expressed as
\begin{equation}\label{VNcond2}
    \sum\limits_{m = 1}^M \delta_m^2 \theta_m^2 < \epsilon^2
\end{equation}
where $\theta_m$ represents the entries of the main diagonal of $\mathbf{L}\mathbf{L}^H$.

Given that $\boldsymbol{\Phi}(\mathbf{b}) = \boldsymbol{\Delta}(\mathbf{b})\mathbf{S}$, we have:
\begin{IEEEeqnarray}{rCl}\label{refeq1}
    \boldsymbol{\Phi}(\mathbf{b}^{(k+1)}) &=& -2Y_0   (\mathbf{Z}_{SS}+r_0\mathbf{I}_M+j\mathbf{B}^{(k+1)})^{-1}\mathbf{S} - \eta_{ct}\frac{Y_0}{Z_0}\mathbf{S} \nonumber\\
    &\approx& -2Y_0 \left(\tilde{\mathbf{A}}^{(k)} - \mathbf{G}^{(k)}\text{diag}\left(\boldsymbol{\delta}\right)\mathbf{A}^{(k)}\right) \mathbf{S} 
\end{IEEEeqnarray}
where $\eta_{ct}$ takes the value 0 for the MP model and 1 for the CT model, so that \eqref{refeq1} is valid for both  models (see \eqref{eq:TFZ_CT1bis}).
Introducing $\mathbf{D}^{(k)} = \mathbf{A}^{(k)}\mathbf{S}$,  $\tilde{\mathbf{D}}^{(k)} = \left(\mathbf{A}^{(k)} + \eta_{ct}\frac{1}{2Z_0}\mathbf{I}_M\right)\mathbf{S}$, from \eqref{notations} and \eqref{refeq1} we obtain
\begin{IEEEeqnarray}{rCl}\label{notations4}
    Q_2 &\approx& 4 Y_0^2 \mathbf{v}^H \bigg\{\left(\tilde{\mathbf{D}}^{(k)} - \mathbf{G}^{(k)}\text{diag}\left(\boldsymbol{\delta}\right)\mathbf{D}^{(k)}\right)^H \mathbf{R}_{{\mathbf{t}_W}} \nonumber\\ 
    && \cdot \left(\tilde{\mathbf{D}}^{(k)} - \mathbf{G}^{(k)}\text{diag}\left(\boldsymbol{\delta}\right)\mathbf{D}^{(k)}\right)\bigg\} \mathbf{v}  \\
    \mathbf{q}_2 &\approx& -2Y_0 \mathbf{T}_{\mathbf{t}_B}  \left(\tilde{\mathbf{D}}^{(k)} - \mathbf{G}^{(k)}\text{diag}\left(\boldsymbol{\delta}\right)\mathbf{D}^{(k)}\right)\mathbf{v}.
\end{IEEEeqnarray}

Introduce now the terms:
\begin{IEEEeqnarray}{rCl}\label{notations4bis}
    C_1 &=& 4 Y_0^2  \mathbf{v}^H\left[\tilde{\mathbf{D}}^{(k)}\right]^H \mathbf{R}_{{\mathbf{t}_W}}\tilde{\mathbf{D}}^{(k)}\mathbf{v} \\
    \mathbf{d} &=& -2 Y_0 \mathbf{T}_{\mathbf{t}_B} \tilde{\mathbf{D}}^{(k)} \mathbf{v} \\
    \mathbf{L}^{(k)}(\boldsymbol{\delta}) &=& \mathbf{G}^{(k)}\text{diag}\left(\boldsymbol{\delta}\right)\mathbf{D}^{(k)}
\end{IEEEeqnarray}
so we can reformulate the expression of $Q_2$ and $\mathbf{q}_2$ as
\begin{IEEEeqnarray}{rCl}\label{notations5}
    Q_2 &\approx&  C_1 -2 \Re \left\{4 Y_0^2 \mathbf{v}^H \left(\tilde{\mathbf{D}}^{(k)}\right)^H\mathbf{R}_{{\mathbf{t}_W}}  \mathbf{L}^{(k)}(\boldsymbol{\delta})   \mathbf{v}\right\}  \\
    && + 4 Y_0^2 \mathbf{v}^H \left[\mathbf{L}^{(k)}(\boldsymbol{\delta})\mathbf{v}\right]^H \mathbf{R}_{{\mathbf{t}_W}}\mathbf{L}^{(k)}(\boldsymbol{\delta})\mathbf{v}\nonumber \\
    \mathbf{q}_2 &\approx& \mathbf{d} + 2Y_0  \mathbf{T}_{\mathbf{t}_B} \mathbf{L}^{(k)}(\boldsymbol{\delta}) \mathbf{v} \nonumber
\end{IEEEeqnarray}
It is now convenient to introduce the terms:
\begin{equation}\label{notations6}
    \begin{array}{ll}
        \mathbf{f}_1 = \mathbf{v}^H \left(\mathbf{D}^{(k)}\right)^H ~,~
        \tilde{\mathbf{f}}_1 = \mathbf{v}^H \left(\tilde{\mathbf{D}}^{(k)}\right)^H  &\in \mathbb{C}^{1 \times M} \\
        \mathbf{F}_1 = \text{diag}(\mathbf{f}_1) ~,~ \tilde{\mathbf{F}}_1 = \text{diag}(\tilde{\mathbf{f}}_1) &\in \mathbb{C}^{M \times M} \\
        \mathbf{F}_2 = \left(\mathbf{G}^{(k)}\right)^H \mathbf{R}_{{\mathbf{t}_W}} \mathbf{G}^{(k)} &\in \mathbb{C}^{M \times M} \\
        \mathbf{F}_3 = \mathbf{R}_{{\mathbf{t}_W}} \mathbf{G}^{(k)} \mathbf{F}^H_1 & \in \mathbb{C}^{M \times M} \\
        \mathbf{F}_4 = \mathbf{T}_{\mathbf{t}_B} \mathbf{G}^{(k)} \mathbf{F}^H_1 & \in \mathbb{C}^{M \times M}
    \end{array}
\end{equation}
Accordingly, by considering only the terms in \eqref{MSE_2} that depend on the optimization variable $\boldsymbol{\delta}$ and exploiting the Neumann approximation, we can approximate the objective function in \eqref{MSE_2} as
\begin{IEEEeqnarray}{rl}
    \hat{\mathcal{G}} & \left(\boldsymbol{\lambda}^{(k+1)},\boldsymbol{\delta}, \mathbf{v} \right) = -4Y_0^2\left \| \boldsymbol{\lambda}^{(k+1)} \right \|^2 \boldsymbol{\delta} \tilde{\mathbf{F}}_1 \mathbf{F}_2   \mathbf{F}^H_1 \boldsymbol{\delta}^T \nonumber\\
    & + 4Y_0 \left[  \Re\left( 2Y_0 \left \|\boldsymbol{\lambda}^{(k+1)} \right \|^2 \tilde{\mathbf{f}}_1 \mathbf{F}_3 + \boldsymbol{\lambda}_2 \mathbf{F}_4  \right)\right]  \boldsymbol{\delta}^T \label{MSE_3}
\end{IEEEeqnarray}

We are now in a position to define the problem for finding the optimal $\boldsymbol{\delta}$ as
\begin{IEEEeqnarray}{rCl}
    \boldsymbol{\delta}^* &=& \arg \min \limits_{\boldsymbol{\delta}} -\hat{\mathcal{G}}\left(\boldsymbol{\lambda}^{(k+1)},\boldsymbol{\delta}\right),\label{Problem3}\\
    \IEEEyessubnumber \textrm{s.t.:} && \sum\limits_{m = 1}^M \delta_m^2 \theta_m^2 < \epsilon^2.
\end{IEEEeqnarray}

The problem in \eqref{Problem3} is now convex and can be easily solved in the Lagrangian domain.
First, from \eqref{MSE_3} we derive the gradient of the Lagrangian as
\begin{IEEEeqnarray}{rCl}
    \nabla_{\boldsymbol{\delta}} &\Bigg[&  -\hat{\mathcal{G}}\left(\boldsymbol{\lambda}^{(k+1)},\boldsymbol{\delta}, \mathbf{v} \right) + \mu \left(\sum\limits_{m = 1}^M \delta_m^2 \theta_m^2-\epsilon^2\right)\Bigg] \nonumber\\
    &=& \boldsymbol{\delta} \left[8Y_0^2  \left \| \boldsymbol{\lambda}^{(k+1)} \right \|^2  \Re\left( \tilde{\mathbf{F}}_1 \mathbf{F}_2   \mathbf{F}^H_1\right)\right]  + 2 \boldsymbol{\delta} \boldsymbol{\Psi} \nonumber\\
    &&- 4Y_0  \Re\left( 2Y_0 \left \|\boldsymbol{\lambda}^{(k+1)} \right \|^2 \tilde{\mathbf{f}}_1 \mathbf{F}_3 + \boldsymbol{\lambda}_2 \mathbf{F}_4  \right)
\end{IEEEeqnarray}
where $\mu > 0$ is a Lagrange multiplier and the vector $\boldsymbol{\Psi} \in \mathbb{C}^{M \times 1}$ has entries $\theta_m^2$.
Accordingly, denoting by
\begin{equation}
    \begin{array}{rcl}
        \mathbf{M} &=& 4Y_0^2  \left \| \boldsymbol{\lambda}^{(k+1)} \right \|^2  \Re\left( \tilde{\mathbf{F}}_1 \mathbf{F}_2   \mathbf{F}^H_1\right)  \\
        \mathbf{p} &=& - 2Y_0  \Re\left( 2Y_0 \left \|\boldsymbol{\lambda}^{(k+1)} \right \|^2 \tilde{\mathbf{f}}_1 \mathbf{F}_3 + \boldsymbol{\lambda}_2 \mathbf{F}_4  \right)
    \end{array}
\end{equation}
we finally obtain
\begin{equation} \label{opt_2.311}
    \boldsymbol{\delta}^* = \left[\mathbf{M}+ \mu \cdot \text{diag}\left(\boldsymbol{\Psi}\right)\right]^{-1}\mathbf{p}
\end{equation}
where $\mu$ is set to satisfy the constraint in \eqref{Problem3}. Finally, we can evaluate $\mathbf{B}^{(k+1)} \approx \mathbf{B}^{(k)} +\mathbf{F}^{(k)}\text{diag}\left(\boldsymbol{\delta}^*\right)$.

\subsection{Precoder optimization}\label{sec:VOptimization}

For the computation of the precoder $\mathbf{v}^{(k+1)}$, another step of AO is performed. Specifically, with the RIS tunable vector fixed to $\mathbf{b}^{(k+1)}$, which for simplicity of notation is denoted as $\mathbf{b}$, we have:
\begin{IEEEeqnarray}{rCl}
    \mathbf{v}^{(k+1)} &=& \arg \min_{\mathbf{v}} -\mathcal{G}\left(\boldsymbol{\lambda}^{(k+1)}, \mathbf{b}, \mathbf{v} \right)\label{opt_2_3bis}\\
    \IEEEyessubnumber \textrm{s.t.:} && \mathbf{v}^H\mathbf{v} \le 1.
\end{IEEEeqnarray}

From \eqref{MSE_2} and \eqref{notations}, introducing 
\begin{equation}\label{notations6bis}
    \begin{array}{rcl}
         \mathbf{W} & = & \mathbf{R}_{{\mathbf{h}_B}} +  \mathbf{R}_{{\mathbf{h}_W}} + \boldsymbol{\Phi}^H(\mathbf{b}) \left(\mathbf{R}_{{\mathbf{t}_B}} + \mathbf{R}_{{\mathbf{t}_W}}\right)\boldsymbol{\Phi}(\mathbf{b})\\
         \mathbf{w}  & = & \boldsymbol{\lambda}^{(k+1)}_1\mathbf{T}_{\mathbf{h}_B} +\boldsymbol{\lambda}^{(k+1)}_2\mathbf{T}_{\mathbf{t}_B} \mathbf{\Phi}(\mathbf{b}) 
    \end{array}
\end{equation}
with $\mathbf{W}\in \mathbb{C}^{N \times N}$ and $\mathbf{w}\in \mathbb{C}^{1 \times N}$, we get
\begin{equation}\label{opt_2_4bis}
    \mathcal{G}\left(\boldsymbol{\lambda}^{(k+1)}, \mathbf{b}, \mathbf{v} \right) =  2\Re\left(\mathbf{w}\mathbf{v}\right)-\mathbf{v}^H \left \| \boldsymbol{\lambda}^{(k+1)} \right \|^2 \mathbf{W}_1 \mathbf{v}.
\end{equation}
The problem in \eqref{opt_2_3bis} is convex and can be solved in the Lagrangian domain. From \eqref{opt_2_4bis} we derive the gradient of the Lagrangian as
\begin{IEEEeqnarray}{rl}
    \nabla_{\mathbf{v}} & \left[-\mathcal{G}\left(\boldsymbol{\lambda}^{(k+1)}, \mathbf{b}, \mathbf{v} \right) + \mu \left(\mathbf{v}^H\mathbf{v} - 1\right)\right]  \nonumber\\
    &= 2\left \| \boldsymbol{\lambda}^{(k+1)} \right \|^2\mathbf{W}\mathbf{v}-2\mathbf{w}^H+2\mu \mathbf{v} \label{gradient11bis}
\end{IEEEeqnarray}
where $\mu > 0$ is a Lagrange multiplier.
Accordingly, we obtain
\begin{equation} \label{opt_2.311bis}
    \mathbf{v}^{(k+1)} = \left[\left \| \boldsymbol{\lambda}^{(k+1)} \right \|^2\mathbf{W}+ \mu \times \mathbf{I}_N\right]^{-1}\mathbf{w}^H
\end{equation} 
where $\mu$ is set to satisfy the constraint in \eqref{opt_2_3bis}. 
Note that the AO approach guarantees that at each iteration:
\begin{IEEEeqnarray}{rl}
\mathcal{G}&\left(\boldsymbol{\lambda}^{(k+2)}, \mathbf{b}^{(k+1)}, \mathbf{v}^{(k+1)} \right) \mathop  \ge \limits^{\left( a \right)} \mathcal{G}\left(\boldsymbol{\lambda}^{(k+1)}, \mathbf{b}^{(k+1)}, \mathbf{v}^{(k+1)} \right) \nonumber\\
&\mathop  \ge \limits^{\left( b \right)} \mathcal{G}\left(\boldsymbol{\lambda}^{(k+1)}, \mathbf{b}^{(k+1)}, \mathbf{v}^{(k)} \right) \mathop  \ge \limits^{\left( c \right)} \mathcal{G}\left(\boldsymbol{\lambda}^{(k+1)}, \mathbf{b}^{(k)}, \mathbf{v}^{(k)}\right) \label{cond1}
\end{IEEEeqnarray}
where $(a)$ follows from $\boldsymbol{\lambda}^{(k+2)} = \boldsymbol{\lambda}^{*}(\mathbf{b}^{(k+1)},\mathbf{v}^{(k+1)})$ where $\boldsymbol{\lambda}^*$ is the minimum defined in \eqref{opt_2_10}, $(b)$, and $(c)$ follow from \eqref{opt_2_3bis} and  \eqref{opt_2_6}, respectively.
From \eqref{cond1} and \eqref{opt_2_1mm}, then we have
\begin{equation}\label{opt_2_2mm}
    \frac{{P}_B(\mathbf{\Phi}(\mathbf{b}^{(k+1)}),\mathbf{v}^{(k+1} )}{{P}_W(\mathbf{\Phi}(\mathbf{b}^{(k+1)}),\mathbf{v}^{(k+1})} \ge \frac{{P}_B(\mathbf{\Phi}(\mathbf{b}^{(k)}),\mathbf{v}^{(k})}{{P}_W(\mathbf{\Phi}(\mathbf{b}^{(k)}),\mathbf{v}^{(k})},
\end{equation}
i.e., the objective function in \eqref{opt_2_3} increases at each iteration. The complete algorithm for the proposed is
summarized in Algorithm \ref{Algo1}. 

\begin{algorithm}[t!]
\footnotesize
\caption{Proposed Optimization Algorithm}
\textbf{Input}: $\mathbf{S}$, $\mathbf{R}_{\mathbf{t}_Q}$, $\mathbf{R}_{\mathbf{h}_Q}$, $\mathbf{T}_{\mathbf{t}_Q}$, $\mathbf{T}_{\mathbf{h}_Q}$ ($Q = B,W$),  $\mathbf{Z}_{SS}$, $Z_0 =50 \, \Omega$, $r_0 > 0$, $\epsilon > 0$\;
\textbf{Initialize}:\\
Generate the initial values of $\mathbf{v}^{(0)}$, $\mathbf{b}^{(0)}$;\\ 
Set an arbitrarily small value $\eta > 0$, $\rho > \eta$, $k \leftarrow 0$\;
\While{$\rho >\eta $}
{
$\mathbf{b} \leftarrow \mathbf{b}^{(k)}$\;
$\mathbf{v} \leftarrow \mathbf{v}^{(k)}$\;
$\mathbf{\Phi} = \mathbf{S}\boldsymbol{\Delta}\left(\mathbf{b}\right)$;\\
$\mathbf{Y}_{B} = \left[\mathbf{T}^T_{\mathbf{h}_B} ,\left(\mathbf{T}_{\mathbf{t}_B}\mathbf{\Phi}\right)^T\right]^T$\;
${P}_W = \mathbf{v}^H \left(\mathbf{R}_{{\mathbf{h}_W}}+\boldsymbol{\Phi}^H \mathbf{R}_{{\mathbf{t}_W}} \boldsymbol{\Phi}\right) \mathbf{v}$\;
$\boldsymbol{\lambda}^{(k+1)} = \frac{\mathbf{v}^H \mathbf{Y}^H_{B}}{ {P}_W}$\;
	Compute $\boldsymbol{\delta}$ according to \eqref{opt_2.311}\;
 $\mathbf{F} ~:~\mathbf{F}_{m,m} =  -\frac{b_m^2+Z_0^2}{2Z_0}$\;
 $\mathbf{B} = \text{diag}(\mathbf{b})$\;
 $\mathbf{B} = \mathbf{B} +\mathbf{F}^{(k)}\text{diag}\left(\boldsymbol{\delta}\right)$\;
 $\mathbf{b} = \text{diag}(\mathbf{B})$\;
	   Compute $\mathbf{v}^{(k+1)}$ according to \eqref{opt_2.311bis}\;
     $\mathbf{b}^{(k+1)} \leftarrow \mathbf{b}$\;
	   $\rho \leftarrow \lVert \mathbf{b}^{(k+1)} - \mathbf{b}^{(k)}\rVert$\;
	   $k \leftarrow k + 1$.
} 
\label{Algo1} 
\end{algorithm}

\subsection{Computational Complexity}
\label{sec:computational_complexity}

The complexity of Algorithm \ref{Algo1} is influenced by the computational effort required for each individual iteration, as well as the total number of iterations needed for convergence, which we discuss in the Results Section. For the evaluation of each iteration, we operate under the assumption that the number of RIS elements significantly exceeds the number of antennas at the base station, specifically $M \gg N$. Furthermore, we consider that the inverse of a matrix involves a computational cost proportional to the cube of its dimensions. 
Given this, it is easy to see that the complexity is heavily influenced by the complexity of computing $\mathbf{b}^{(k+1)}$, which is $\mathcal{O}(M^3)$, stemming from the evaluation of $\mathbf{A}^{(k)}$ in \eqref{notations3A} and the matrix inversion in \eqref{opt_2.311}. Therefore, considering only the most significant terms, the complexity of each iteration of Algorithm \ref{Algo1} is $\mathcal{O}(M^3)$. Thus, the total complexity turns out to be $\mathcal{O}(N_{\text{it}} M^3)$, with $N_{\text{it}}$ being the number of iterations of the algorithm.
In particular, Sec.~\ref{sec:NumericalResults} shows that convergence is generally achieved in a few hundred iterations.

Although a direct comparison with other approaches is difficult to make, since the assumption of knowing second-order statistics and the use of an MP model for the RIS are peculiar characteristics of this work, it can be noted that, in general, the problem of optimizing the RIS for similar scenarios (e.g., \cite{CovertCommRIS,DelayContraintRISCC}) is addressed in the literature using the Semi-Definite Relaxation (SDR) approach, with a complexity on the order of $\mathcal{O}(M^{6.5})$ per iteration.
Therefore, the proposed approach reduces significantly the complexity compared to conventional approaches proposed for similar problems. 

\section{Numerical Results} \label{sec:NumericalResults}
In this section, the performance of the proposed method (exploiting second-order CSI, sCSI), where the RIS is modeled using the MP model (sCSI-MP), is evaluated in four different scenarios corresponding to various positions of Alice, Bob, and Willie. To highlight the importance of prior knowledge of Willie's position uncertainty region, and therefore of the second-order channel statistics reported in \eqref{eq:RtQ} and \eqref{eq:RhQ}, we consider as a baseline solution (SoA-MP) the one in which the second-order channel statistics are unknown.
In this case, as in \cite{CovertCommRIS,WardenStatCsiRISCC,DelayContraintRISCC}, the channels towards Willie are assumed to be spatially i.i.d. Rayleigh fading channels for the purposes of optimizing the RIS and the precoder. Subsequently, to meet the covert constraint expressed in \eqref{eq:CovertConstraint}, the transmitted power is scaled accordingly, as in the sCSI-MP case.
Moreover, we present the performance of the proposed method when, for RIS and precoder optimization, the RIS is modeled using the CT model (sCSI-CT), while the results are evaluated considering the more comprehensive MP model. This allows us to appreciate the extent of the mismatch due to the use of the approximate CT model. Finally, we also consider the case where the RIS is absent from the system (w/o-RIS), so that only direct links are available, and the optimization focuses solely on Alice's precoder.

\begin{table}[t]
\centering
\begin{tabular}{|l|c|c|c|c|c|c|}
\hline
\rule[-4pt]{0pt}{13pt}& $\vartheta_A$ & $\overline{\vartheta}_B$ & $\overline{\vartheta}_W$ & $d_A$ & $d_W$ & $\Delta \phi_W$ \\ \hline
\rule[-5pt]{0pt}{14pt}Scenario 1 & $\frac{3}{4}\pi$ & $\frac{1}{8}\pi$ & $\frac{3}{8}\pi$ & $30~\meter$ & $30~\meter$ & $\frac{1}{4}\pi$\\ \hline
\rule[-5pt]{0pt}{14pt}Scenario 2 & $\frac{1}{2}\pi$ & $\frac{1}{8}\pi$ & $\frac{1}{4}\pi$ & $30~\meter$ & $30~\meter$ & $\frac{1}{8}\pi$\\ \hline
\rule[-5pt]{0pt}{14pt}Scenario 3 & $\frac{3}{4}\pi$ & $\frac{7}{16}\pi$ & $\frac{1}{8}\pi$ & $50~\meter$ & $30~\meter$ & $\frac{1}{4}\pi$\\ \hline
\rule[-5pt]{0pt}{14pt}Scenario 4 & $\frac{3}{4}\pi$ & $\frac{1}{8}\pi$ & $\frac{1}{4}\pi$ & $40~\meter$ & $21~\meter$ & $\frac{1}{16}\pi$\\ \hline
\end{tabular}
\caption{Simulation parameters for the different scenarios}
\label{tab:Params}
\end{table}

The considered scenario is the one depicted in Fig.~\ref{Fig:CoverCommRIS}, with carrier frequency set to $f_0 = 3.6$~\giga\hertz, corresponding to the highest frequency in the 5G standard for sub-$6$~\giga\hertz~transmissions.
The RIS is configured as a uniform planar array (UPA) with $M_H = 32$ elements along the $x$ axis and $M_V = 4$ elements along the $z$ axis. As in \cite{abrardo_E}, it is  considered that the RIS elements  are modeled as identical $z$-oriented metallic thin wire dipoles with a radius $\lambda/500$ and length $0.46 \lambda$. The dipole length  has been chosen to have nearly resonant elements characterized by a low reactance of the input impedance. The spacing between the RIS dipoles is set to $d_x = \lambda/2$ in the $x$ direction and to $d_z = 3/4 \lambda$ in the $z$ direction, so that the dimension of the RIS is approximately $1.33 \times 0.5$ meters. 
Alice is equipped with $N = 16$ dipole antennas, arranged in $2$ linear arrays along the $z$ axis, with each array containing $8$ elements along the $x$ axis. Hence, Bob and Willie are equipped with single dipole antennas. The arrangement and type of antennas at Alice, Bob, and Willie are the same as those of the RIS described above.
The parameters reported in the RIS model of \eqref{eq:TFZ} are set according to the analytical approach presented in \cite{GradoniEnd2End}.

Regarding the channel characterization, we consider a scenario in which the RIS is positioned at an elevated and advantageous location, so that the direct and reflected channels are modeled differently. 
For the direct path, a typical urban scenario is considered, where there is a strong NLOS component due to environmental scattering. Accordingly, a pathloss exponent of 3.5, typical of urban environments, and a Rice factor $K=1$ are considered. For the paths involving the RIS, however, a strong LOS component and a small NLOS component are assumed, reflecting the expectation that a link with an elevated RIS experiences fewer impairments. Specifically, $K=10$ is considered with a pathloss exponent set to $2$. 
Moreover, for the same reasons, the scattering angular spread $\Delta_m$ is considered to be equal to $\frac{\pi}{6}$ and $\frac{\pi}{64}$ for the direct and reflected links, respectively.

The simulation results are obtained for four different scenarios, corresponding to four different positions of the nodes.
The positions are described in terms of polar coordinates on the $xy$ plane with a specific height along the $z$ axis.
Specifically, the RIS is positioned at the origin of the $xy$ plane at a height of $5~\meter$, while the nodes are characterized by a height of $3$ meters.
Regarding the positions of Willie and Bob on the $xy$ plane, in line with the approach described in Section \ref{sec:SystemModel}, they are located within certain regions $\mathcal{P}_W$ and $\mathcal{P}_B$, with corresponding distributions $f{\boldsymbol{\wp}_W}(\mathbf{p})$ and $f{\boldsymbol{\wp}_B}(\mathbf{p})$. The correlations in \eqref{eq:RtQ} and \eqref{eq:RhQ} are then calculated using numerical integration over the respective areas, and the results are then obtained through a Monte Carlo simulation, averaging the quantities of interest over all possible positions of the nodes.
To facilitate this calculation without losing generality, it is assumed that the uncertainty regions are one-dimensional and, in particular, are characterized by a certain angular uncertainty with uniform distribution and with a fixed distance. 
Accordingly, Willie and Bob's positions are considered to be in an arc with respect to the RIS, where the aperture angle is related to the position uncertainty.
While a fixed uncertainty angle of $\frac{\pi}{64}$ is considered for Bob\footnote{We assumed a small uncertainty angle for Bob as he is part of Alice's network, so the position uncertainty is mainly related to measurement accuracy}, the uncertainty angle for Willie is specified in different scenarios and is generally wider.

Therefore, for each scenario the specific parameters are:
\begin{itemize}
    \item the angular position of Alice $\vartheta_A$, and the average angular position of Bob and Willie, $\overline{\vartheta}_B$ and $\overline{\vartheta}_W$, respectively;
    \item Alice and Willie distance from the RIS, $d_A$ and $d_W$;
    \item Willie uncertainty angle, $\Delta \phi_W$.
\end{itemize}
These parameters are summarized in Tab.~\ref{tab:Params} for the different scenarios.
In the following figures, results are provided in terms of Bob's achievable rate versus the complementary minimum DEP ($\delta=1-\zeta_{min}$).

\begin{figure}
    \centering
    \includegraphics{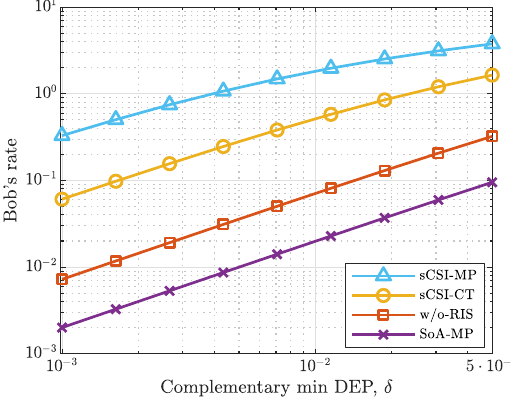}
    \caption{Performance comparison in the first scenario.}
    \label{Fig:Rb_vs_Zeta_scenario_1}
\end{figure}

\begin{figure}
    \centering
    \includegraphics{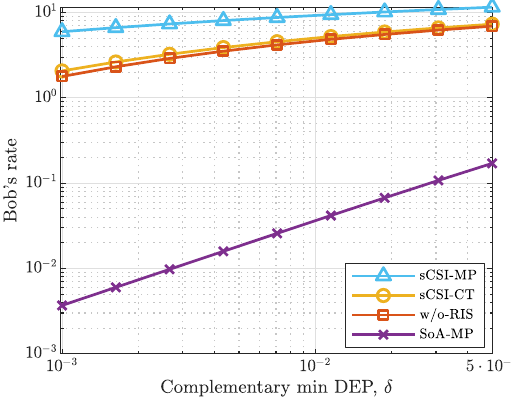}
    \caption{Performance comparison in the second scenario.}
    \label{Fig:Rb_vs_Zeta_scenario_2}
\end{figure}

In the first scenario, Willie and Bob are positioned symmetrically with respect to Alice's main antenna axis (the $x$ axis), such that any beam directed toward Bob also generates a beam directed toward Willie's area.
Under these conditions, the direct channel cannot produce good results, and it is natural to expect that the optimizer will choose to use only the path reflected by the RIS, as will be shown in a subsequent figure.
The results for all the considered schemes in this scenario are shown in Fig.~\ref{Fig:Rb_vs_Zeta_scenario_1}.
It can be seen that the proposed method (sCSI-MP) achieves the best performance, presenting the highest rate.
On the other hand, applying the same optimization approach with the CT RIS model (sCSI-CT) shows some performance degradation, highlighting that the mismatch in the RIS model is a non-negligible factor.
Moreover, without RIS assistance (w/o-RIS), performance suffers significant degradation as the optimizer is forced to rely solely on the direct link.
However, even in this scenario, the performance is better than in the baseline (SoA-MP). This occurs because, in the SoA-MP approach, the RIS is optimized without knowing the statistical position of Willie, and thus it is effectively not utilized, as the direct path is stronger. Hence, due to the interference generated toward Willie, the need for power scaling to satisfy the covert constraint makes this method the least effective.

In the second scenario, Alice is positioned directly in front of the RIS, with Bob's and Willie's areas located at different angles relative to Alice.
Accordingly, the angular selectivity of Alice's antenna array enables effective use of the direct path.
This is evident in Fig.~\ref{Fig:Rb_vs_Zeta_scenario_2}, where sCSI-MP continues to outperform the cases sCSI-CT and w/o-RIS, though the margin with respect to w/o-RIS is reduced compared to the first scenario.
Indeed, the w/o-RIS case also shows quite good performance, underscoring the ability of Alice's antenna array to focus the beam toward Bob without significant leakage toward Willie.
Finally, the SoA-MP method shows significantly worse performance.

\begin{figure}
    \centering
    \includegraphics{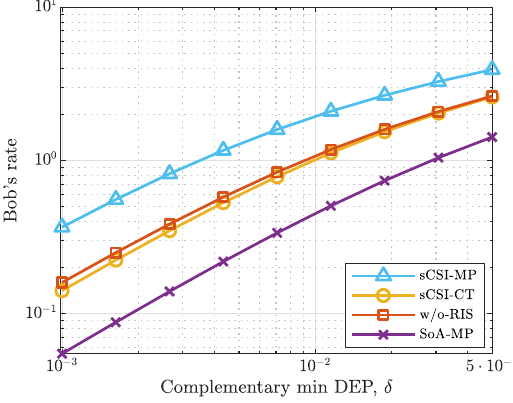}
    \caption{Performance comparison in the third scenario.}
    \label{Fig:Rb_vs_Zeta_scenario_3}
\end{figure}

\begin{figure}
    \centering
    \includegraphics{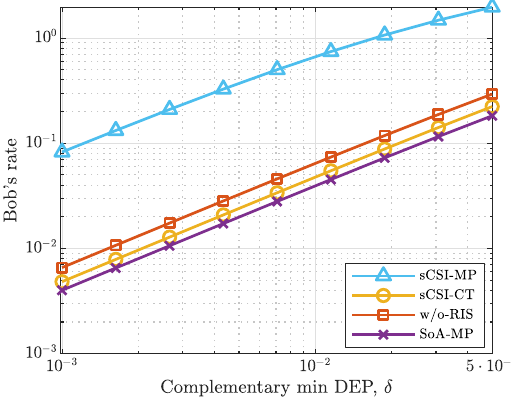}
    \caption{Performance comparison in the fourth scenario.}
    \label{Fig:Rb_vs_Zeta_scenario_4}
\end{figure}

In the third scenario, Alice is positioned farther away ($d_A = 50~\meter$), Willie's uncertainty angle is wider, \textcolor{blue}{and Bob's area is much closer to Alice compared to Willie. In this case, the performance is generally worse than in the previous cases.}
The best scheme continues to be sCSI-MP, which effectively combines the direct and reflected paths, although, as in the previous case, the direct path alone can still be used efficiently.
In fact, from Fig.~\ref{Fig:Rb_vs_Zeta_scenario_3}, it is possible to see, as in the previous case, that the w/o-RIS method has performance very close to sCSI-CT.
In this case, SoA-MP is still the method with the poorest performance, but with a reduced margin, as only the direct path is used and it does not produce significant leakage since Willie is farther from Alice than Bob.

In the fourth scenario, Willie's area is positioned along the direct path between Alice and Bob, rendering the direct link unusable for covert communication.
In this case, it is expected that the no-RIS scheme will perform very poorly, as will the SoA-MP RIS, which, lacking information on possible positions of Willie, will tend to rely heavily on the direct path.
Additionally, Willie is also aligned with the specular component of the path reflected by the RIS, which would generate a significant signal component directed toward him if no countermeasures are taken.
Such interference can be partially compensated in the sCSI-MP case because the RIS model includes this component.
In contrast, in the CT - MP RIS case, this is not possible, as the specular component is not taken into account in the model.
Accordingly, as shown in Fig.~\ref{Fig:Rb_vs_Zeta_scenario_4}, the sCSI-MP approach demonstrates significantly better performance than all the alternatives.

\begin{figure}
    \centering
    \includegraphics{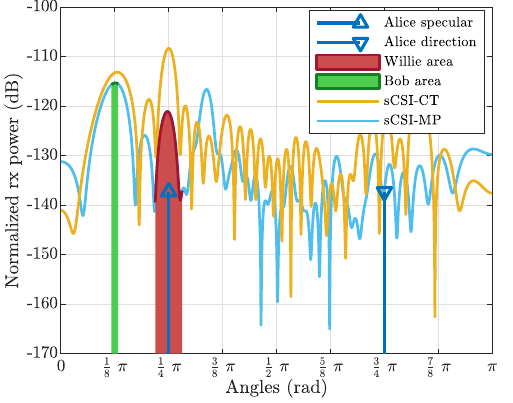}
    \caption{RIS radiation diagram for sCSI-MP in the fourth scenario.}
    \label{Fig:RadDiagram}
\end{figure}

The importance of the RIS MP model in the fourth scenario is underscored in Fig.~\ref{Fig:RadDiagram}, which depicts the RIS radiation pattern for both the sCSI-MP and the sCSI-CT approaches.
Moreover, the diagram also shows the angular areas where Willie and Bob are located, as well as the direction of the signal arriving from Alice (Alice direction) along with the corresponding direction of its specular component (Alice specular). 
It is evident that the optimization based on the CT RIS model directs more power toward Bob but suffers from significant leakage in the specular direction where Willie is positioned.
In contrast, the sCSI-MP approach effectively reduces the specular component, sacrificing a few decibels in Bob's direction.
This trade-off results in far superior performance by minimizing the risk of detection by Willie.

\begin{figure}
    \centering
    \includegraphics{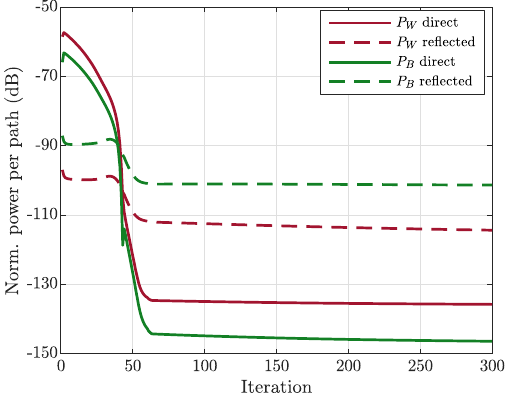}
    \caption{Iteration convergence of sCSI-MP.}
    \label{Fig:Pr_vs_iter_scenario}
\end{figure}

Finally, Fig.~\ref{Fig:Pr_vs_iter_scenario} illustrates the convergence properties of the sCSI-MP approach in the first scenario.
The plot shows the average normalized power received by Bob (in green) and Willie (in red) as a function of the algorithm's iterations, with solid lines representing the direct link and dashed lines representing the link involving the RIS.
Initially, the algorithm allocates most of the power to the direct links. However, after approximately 45 iterations, this trend reverses, and power is gradually shifted away from the direct links.
This indicates that the RIS optimization has reached a configuration that enables enhanced performance.
Note that the average power received from the direct links in the early iterations is higher than that received from the RIS at convergence, which means that the direct link is better in terms of attenuation.
However, due to the spatial arrangements of the nodes, it is beneficial, as discussed earlier, to primarily use the paths reflected by the RIS.
It is noted that the algorithm achieves substantial convergence after approximately $N_{it} = 300$ iterations, and a similar result is obtained for the other scenarios, which are not shown here due to space limitations.


\section{Conclusion}\label{sec:Conclusion}
This paper introduces an innovative optimization strategy for Reconfigurable Intelligent Surfaces (RIS) designed specifically for covert communications.
Our model incorporates often-neglected aspects of conventional RIS models in communication theory, such as mutual coupling between scattering elements and the presence of a specular component in the signal arriving at the RIS.
Furthermore, the proposed approach takes into account the limited channel state information available between the Alice and Bob, and between Alice and Willie.
Numerical simulations demonstrate the effectiveness of our method, highlighting its capability to enable communication with the authorized user while maintaining the required level of covertness.
This novel approach addresses key challenges in RIS-aided covert communications, offering a more comprehensive and realistic model for practical applications.

\bibliographystyle{IEEEtran}
\bibliography{IEEEabrv,RIS_covert_channel_biblio}

\begin{thebibliography}{10}
\providecommand{\url}[1]{#1}
\csname url@samestyle\endcsname
\providecommand{\newblock}{\relax}
\providecommand{\bibinfo}[2]{#2}
\providecommand{\BIBentrySTDinterwordspacing}{\spaceskip=0pt\relax}
\providecommand{\BIBentryALTinterwordstretchfactor}{4}
\providecommand{\BIBentryALTinterwordspacing}{\spaceskip=\fontdimen2\font plus
\BIBentryALTinterwordstretchfactor\fontdimen3\font minus
  \fontdimen4\font\relax}
\providecommand{\BIBforeignlanguage}[2]{{%
\expandafter\ifx\csname l@#1\endcsname\relax
\typeout{** WARNING: IEEEtran.bst: No hyphenation pattern has been}%
\typeout{** loaded for the language `#1'. Using the pattern for}%
\typeout{** the default language instead.}%
\else
\language=\csname l@#1\endcsname
\fi
#2}}
\providecommand{\BIBdecl}{\relax}
\BIBdecl

\bibitem{CovertComm}
B.~A. Bash, D.~Goeckel, D.~Towsley, and S.~Guha, ``{Hiding information in
  noise: fundamental limits of covert wireless communication},'' \emph{{IEEE}
  Commun. Mag.}, vol.~53, no.~12, pp. 26--31, 2015.

\bibitem{angevine2019}
R.~G. Angevine, J.~K. Warden, R.~Keller, and C.~Frye, ``{Learning Lessons from
  the Ukraine Conflict},'' Institute for Defense Analyses, May 2019.

\bibitem{RIS}
E.~Basar, M.~Di~Renzo, J.~De~Rosny, M.~Debbah, M.-S. Alouini, and R.~Zhang,
  ``{Wireless Communications Through Reconfigurable Intelligent Surfaces},''
  \emph{{IEEE} Access}, vol.~7, pp. 116\,753--116\,773, 2019.

\bibitem{CovertCommLimits}
B.~A. Bash, D.~Goeckel, and D.~Towsley, ``{Limits of Reliable Communication
  with Low Probability of Detection on AWGN Channels},'' \emph{{IEEE} J. Sel.
  Areas Commun.}, vol.~31, no.~9, pp. 1921--1930, 2013.

\bibitem{UndetectableComm}
S.~Lee, R.~J. Baxley, M.~A. Weitnauer, and B.~Walkenhorst, ``{Achieving
  Undetectable Communication},'' \emph{{IEEE} J. Sel. Topics Signal Process.},
  vol.~9, no.~7, pp. 1195--1205, 2015.

\bibitem{CovertCommUninformedJammer}
T.~V. Sobers, B.~A. Bash, S.~Guha, D.~Towsley, and D.~Goeckel, ``{Covert
  Communication in the Presence of an Uninformed Jammer},'' \emph{{IEEE} Trans.
  Wireless Commun.}, vol.~16, no.~9, pp. 6193--6206, 2017.

\bibitem{CovertCommNoiseUncertainty}
B.~He, S.~Yan, X.~Zhou, and V.~K.~N. Lau, ``{On Covert Communication With Noise
  Uncertainty},'' \emph{{IEEE} Commun. Lett.}, vol.~21, no.~4, pp. 941--944,
  2017.

\bibitem{CovertCommRelay}
J.~Hu, S.~Yan, X.~Zhou, F.~Shu, J.~Li, and J.~Wang, ``{Covert Communication
  Achieved by a Greedy Relay in Wireless Networks},'' \emph{{IEEE} Trans.
  Wireless Commun.}, vol.~17, no.~7, pp. 4766--4779, 2018.

\bibitem{AI4CovertComm}
X.~Liao, J.~Si, J.~Shi, Z.~Li, and H.~Ding, ``{Generative Adversarial Network
  Assisted Power Allocation for Cooperative Cognitive Covert Communication
  System},'' \emph{{IEEE} Commun. Lett.}, vol.~24, no.~7, pp. 1463--1467, 2020.

\bibitem{SISORIS}
Q.~Wu and R.~Zhang, ``{Towards Smart and Reconfigurable Environment:
  Intelligent Reflecting Surface Aided Wireless Network},'' \emph{{IEEE}
  Commun. Mag.}, vol.~58, no.~1, pp. 106--112, 2020.

\bibitem{MURIS}
H.~Guo, Y.-C. Liang, J.~Chen, and E.~G. Larsson, ``{Weighted Sum-Rate
  Maximization for Reconfigurable Intelligent Surface Aided Wireless
  Networks},'' \emph{{IEEE} Trans. Wireless Commun.}, vol.~19, no.~5, pp.
  3064--3076, 2020.

\bibitem{EnergyEfficiencyRIS}
C.~Huang, A.~Zappone, G.~C. Alexandropoulos, M.~Debbah, and C.~Yuen,
  ``{Reconfigurable Intelligent Surfaces for Energy Efficiency in Wireless
  Communication},'' \emph{{IEEE} Trans. Wireless Commun.}, vol.~18, no.~8, pp.
  4157--4170, 2019.

\bibitem{LocalizationRIS}
S.~Palmucci, A.~Guerra, A.~Abrardo, and D.~Dardari, ``{Two-Timescale Joint
  Precoding Design and RIS Optimization for User Tracking in Near-Field MIMO
  Systems},'' \emph{{IEEE} Trans. Signal Process.}, vol.~71, pp. 3067--3082,
  2023.

\bibitem{SecrecyRIS}
M.~Cui, G.~Zhang, and R.~Zhang, ``{Secure Wireless Communication via
  Intelligent Reflecting Surface},'' \emph{{IEEE} Wireless Commun. Lett.},
  vol.~8, no.~5, pp. 1410--1414, 2019.

\bibitem{RisScatteringParameterNetworkAnalysis}
S.~Shen, B.~Clerckx, and R.~Murch, ``{Modeling and Architecture Design of
  Reconfigurable Intelligent Surfaces Using Scattering Parameter Network
  Analysis},'' \emph{{IEEE} Trans. Wireless Commun.}, vol.~21, no.~2, pp.
  1229--1243, 2022.

\bibitem{abrardo_E}
A.~Abrardo, A.~Toccafondi, and M.~D. Renzo, ``{Design of Reconfigurable
  Intelligent Surfaces by Using S-Parameter Multiport Network Theory –
  Optimization and Full-Wave Validation},'' \emph{{IEEE} Trans. Wireless
  Commun.}, pp. 1--1, 2024.

\bibitem{GradoniEnd2End}
G.~Gradoni and M.~Di~Renzo, ``{End-to-End Mutual Coupling Aware Communication
  Model for Reconfigurable Intelligent Surfaces: An Electromagnetic-Compliant
  Approach Based on Mutual Impedances},'' \emph{{IEEE} Wireless Commun. Lett.},
  vol.~10, no.~5, pp. 938--942, 2021.

\bibitem{DR2}
X.~Qian and M.~D. Renzo, ``{Mutual Coupling and Unit Cell Aware Optimization
  for Reconfigurable Intelligent Surfaces},'' \emph{{IEEE} Wireless Commun.
  Lett.}, vol.~10, no.~6, pp. 1183--1187, 2021.

\bibitem{ABR_MUTUAL}
A.~Abrardo, D.~Dardari, M.~Di~Renzo, and X.~Qian, ``{MIMO Interference Channels
  Assisted by Reconfigurable Intelligent Surfaces: Mutual Coupling Aware
  Sum-Rate Optimization Based on a Mutual Impedance Channel Model},''
  \emph{{IEEE} Wireless Commun. Lett.}, vol.~10, no.~12, pp. 2624--2628, 2021.

\bibitem{IdeaRISCC}
X.~Lu, E.~Hossain, T.~Shafique, S.~Feng, H.~Jiang, and D.~Niyato,
  ``{Intelligent Reflecting Surface Enabled Covert Communications in Wireless
  Networks},'' \emph{{IEEE} Netw.}, vol.~34, no.~5, pp. 148--155, 2020.

\bibitem{CovertDetectionRIS}
Z.~Chen, S.~Yan, X.~Zhou, F.~Shu, and D.~W.~K. Ng, ``{Intelligent Reflecting
  Surface-Assisted Passive Covert Wireless Detection},'' \emph{{IEEE} Trans.
  Veh. Technol.}, vol.~73, no.~2, pp. 2954--2959, 2024.

\bibitem{MIMORISCC}
X.~Chen, T.-X. Zheng, L.~Dong, M.~Lin, and J.~Yuan, ``{Enhancing MIMO Covert
  Communications via Intelligent Reflecting Surface},'' \emph{{IEEE} Wireless
  Commun. Lett.}, vol.~11, no.~1, pp. 33--37, 2022.

\bibitem{EnergyEfficiencyRISCC1}
Z.~Yang, P.~Yue, S.~Wang, G.~Pan, and J.~An, ``{Energy-Efficient Optimization
  for RIS-Aided MIMO Covert Communications},'' \emph{{IEEE} Internet Things
  J.}, vol.~10, no.~21, pp. 18\,993--19\,003, 2023.

\bibitem{EnergyEfficiencyRISCC2}
M.~Li, X.~Tao, N.~Li, and H.~Wu, ``{Energy-Efficient Covert Communication With
  the Aid of Aerial Reconfigurable Intelligent Surface},'' \emph{{IEEE} Commun.
  Lett.}, vol.~26, no.~9, pp. 2101--2105, 2022.

\bibitem{FriendlyJammerRISCC}
J.~Kong, F.~T. Dagefu, J.~Choi, R.~Aggarwal, and P.~Spasojevic, ``{Covert
  Communication in Intelligent Reflecting Surface Assisted Networks With a
  Friendly Jammer},'' \emph{{IEEE} Trans. Veh. Technol.}, vol.~73, no.~1, pp.
  1467--1472, 2024.

\bibitem{CovertCommRIS}
J.~Si, Z.~Li, Y.~Zhao, J.~Cheng, L.~Guan, J.~Shi, and N.~Al-Dhahir, ``{Covert
  Transmission Assisted by Intelligent Reflecting Surface},'' \emph{{IEEE}
  Trans. Commun.}, vol.~69, no.~8, pp. 5394--5408, 2021.

\bibitem{WardenStatCsiRISCC}
C.~Wu, S.~Yan, X.~Zhou, R.~Chen, and J.~Sun, ``{Intelligent Reflecting Surface
  (IRS)-Aided Covert Communication With Warden’s Statistical CSI},''
  \emph{{IEEE} Wireless Commun. Lett.}, vol.~10, no.~7, pp. 1449--1453, 2021.

\bibitem{DelayContraintRISCC}
X.~Zhou, S.~Yan, Q.~Wu, F.~Shu, and D.~W.~K. Ng, ``{Intelligent Reflecting
  Surface (IRS)-Aided Covert Wireless Communications With Delay Constraint},''
  \emph{{IEEE} Trans. Wireless Commun.}, vol.~21, no.~1, pp. 532--547, 2022.

\bibitem{ActivePassiveBeamFormingRISCC}
C.~Wang, Z.~Li, J.~Shi, and D.~W.~K. Ng, ``{Intelligent Reflecting
  Surface-Assisted Multi-Antenna Covert Communications: Joint Active and
  Passive Beamforming Optimization},'' \emph{{IEEE} Trans. Wireless Commun.},
  vol.~69, no.~6, pp. 3984--4000, 2021.

\bibitem{1499046}
M.~McKay and I.~Collings, ``{General capacity bounds for spatially correlated
  Rician MIMO channels},'' \emph{{IEEE} Trans. Inf. Theory}, vol.~51, no.~9,
  pp. 3121--3145, 2005.

\bibitem{9328501}
X.~Wei, D.~Shen, and L.~Dai, ``{Channel Estimation for RIS Assisted Wireless
  Communications Part I: Fundamentals, Solutions, and Future Opportunities},''
  \emph{{IEEE} Commun. Lett.}, vol.~25, no.~5, pp. 1398--1402, 2021.

\bibitem{Demir2022is}
{\"O}.~T. Demir and E.~Bj{\"o}rnson, ``{Is Channel Estimation Necessary to
  Select Phase-Shifts for RIS-Assisted Massive MIMO?}'' \emph{{IEEE} Trans.
  Wireless Commun.}, vol.~21, no.~11, pp. 9537--9552, 2022.

\bibitem{ABR}
A.~Abrardo, D.~Dardari, and M.~Di~Renzo, ``{Intelligent Reflecting Surfaces:
  Sum-Rate Optimization Based on Statistical Position Information},''
  \emph{{IEEE} Trans. Commun.}, vol.~69, no.~10, pp. 7121--7136, 2021.

\bibitem{WuZhang2020}
M.-M. Zhao, Q.~Wu, M.-J. Zhao, and R.~Zhang, ``{Intelligent Reflecting Surface
  Enhanced Wireless Networks: Two-Timescale Beamforming Optimization},''
  \emph{{IEEE} Trans. Wireless Commun.}, vol.~20, no.~1, pp. 2--17, 2021.

\bibitem{Abrardo_two}
F.~Jiang, A.~Abrardo, K.~Keykhosravi, H.~Wymeersch, D.~Dardari, and
  M.~Di~Renzo, ``{Two-Timescale Transmission Design and RIS Optimization for
  Integrated Localization and Communications},'' \emph{{IEEE} Trans. Wireless
  Commun.}, vol.~22, no.~12, pp. 8587--8602, 2023.

\bibitem{Demir2022Channel}
{\"O}.~T. Demir, E.~Bj{\"o}rnson, and L.~Sanguinetti, ``Channel modeling and
  channel estimation for holographic massive mimo with planar arrays,''
  \emph{{IEEE} Wireless Commun. Lett.}, vol.~11, no.~5, pp. 997--1001, 2022.

\bibitem{Bjornson2021Rayleigh}
E.~Bj{\"o}rnson and L.~Sanguinetti, ``{Rayleigh Fading Modeling and Channel
  Hardening for Reconfigurable Intelligent Surfaces},'' \emph{{IEEE} Wireless
  Commun. Lett.}, vol.~10, no.~4, pp. 830--834, 2021.

\bibitem{RandomWirelessNetworks}
T.-X. Zheng, H.-M. Wang, D.~W.~K. Ng, and J.~Yuan, ``{Multi-Antenna Covert
  Communications in Random Wireless Networks},'' \emph{{IEEE} Trans. Wireless
  Commun.}, vol.~18, no.~3, pp. 1974--1987, 2019.

\bibitem{FractionalProgramming}
K.~Shen and W.~Yu, ``{Fractional Programming for Communication Systems—Part
  I: Power Control and Beamforming},'' \emph{{IEEE} Trans. Signal Process.},
  vol.~66, no.~10, pp. 2616--2630, 2018.

\end{thebibliography}
\end{document}